\documentclass[twoside]{article}
\usepackage{graphicx,amssymb,mathrsfs,amsmath}
\usepackage[numbers,sort&compress]{natbib}
\catcode`@=11
\long\def\@makefntext#1{\noindent #1}
\newskip\tabcentering \tabcentering=1000pt plus 1000pt minus 1000pt
\def\MCH#1#2{\setbox0=\hbox{\raise#1\hbox{#2}}\smash{\box0}}

\def\CR{\cr\noalign{\vspace{1mm} \hrule \vspace{1mm}}}
\def\@evenfoot{}\def\@oddfoot{}

\def\@evenhead{\hbox to\textwidth{\footnotesize\rm\thepage \hfill
{\it Tiexin Guo}}} 

\def\@oddhead{\hbox to \textwidth{\footnotesize{\it
Recent progress in random metric theory and its applications to conditional risk measures } \hfill\thepage}}

\def\sec#1{\vskip 3mm\leftline{\bf #1}\vskip 1mm}
\def\subsec#1{\vskip 2mm\leftline{#1}\vskip 1mm}
\def\th#1{\vskip 1mm\noindent{\bf #1}\quad}

\renewcommand{\topfraction}{1}
\renewcommand{\bottomfraction}{1}
\renewcommand{\textfraction}{0}
\renewcommand{\floatpagefraction}{0}
\catcode`@=12

\def\R{{\Bbb R}}

\def\bc{\begin{center}}
\def\ec{\end{center}}
\def\no{\noindent}
\def\hang{\hangindent\parindent}
\def\textindent#1{\indent\llap{\qquad #1\ \ \enspace}\ignorespaces}
\def\ref{\par\hang\textindent}

\def\dl{\displaystyle\lim}

\begin{document}
\abovedisplayskip=6pt plus 1pt minus 1pt \belowdisplayskip=6pt
plus 1pt minus 1pt
\thispagestyle{empty} \vspace*{-1.0truecm} \noindent
\vskip 10mm

\bc{\Large\bf Recent progress in random metric theory and its applications to conditional risk measures

\footnotetext{\footnotesize $^{\ast}$
E-mail address: txguo@buaa.edu.cn }
\footnotetext{\footnotesize Supported by NNSF No. 10871016}} \ec

\vskip 5mm
\bc{\bf Tiexin Guo$^{\ast}$}\\
{\small LMIB and School of Mathematics and Systems Science,
Beihang University, \\
Beijing 100191, P.R. China.}\ec

\def\sec#1{\vspace{2mm}\noindent{{\bf #1}}\vspace{0.5mm}}
\def\subsec#1{\vspace{2mm}\leftline{\bf #1}} 
\def\th#1{\vspace{1mm}\noindent{\bf #1}\quad } 
\def\pf#1{\vspace{1mm}\noindent{\it #1}\quad}
\renewcommand{\topfraction}{1} \renewcommand{\bottomfraction}{1}
\renewcommand{\textfraction}{0} \renewcommand{\floatpagefraction}{0}
\floatsep=0pt \textfloatsep=0pt \intextsep=0pt \catcode`@=12
\def\leq{\leqslant}
\def\geq{\geqslant}
\def\R{{\Bbb R}}  \def\N{{\Bbb N}}  \def\Q{{\Bbb Q}}  \def\O{{\Bbb O}} \def\Z{{\Bbb Z}}
\def\C{{\Bbb C}}  \def\hml{\end{document}}  \newsymbol\wjzhml 203F \def\no{\noindent}
\def\CR{\cr\noalign{\vspace{1mm} \hrule \vspace{1mm}}}

\abovedisplayskip=3pt plus 1pt minus 1pt 
\belowdisplayskip=3pt plus 1pt minus 1pt 

\def\le{\leqslant}
\def\ge{\geqslant}
\def\dl{\displaystyle}




\vspace{8true mm}

\renewcommand{\baselinestretch}{1.9}\baselineskip 19pt

\baselineskip 12pt \renewcommand{\baselinestretch}{1.18}
\noindent{{\bf Abstract}\small\hspace{2.8mm} 
The purpose of this paper is to give a selective survey on recent
progress in random metric theory and its applications to conditional
risk measures. This paper includes eight sections. Section 1 is a
longer introduction, which gives a brief introduction to random
metric theory, risk measures and conditional risk measures. Section
2 gives the central framework in random metric theory, topological
structures, important examples, the notions of a random conjugate
space and the Hahn-Banach theorems for random linear functionals.
Section 3 gives several important representation theorems for random
conjugate spaces. Section 4 gives characterizations for a complete
random normed module to be random reflexive. Section 5 gives
hyperplane separation theorems currently available in random locally
convex modules. Section 6 gives the theory of random duality with
respect to the locally $L^{0}-$convex topology and in particular a
characterization for a locally $L^{0}-$convex module to be
$L^{0}-$pre$-$barreled. Section 7 gives some basic results on
$L^{0}-$convex analysis together with some applications to
conditional risk measures. Finally, Section 8 is devoted to
extensions of conditional convex risk measures, which shows that
every representable $L^{\infty}-$type of conditional convex risk
measure and every continuous $L^{p}-$type of convex conditional risk
measure ($1\leq p<+\infty$) can be extended to an $L^{\infty}_{\cal
F}({\cal E})-$type of $\sigma_{\epsilon,\lambda}(L^{\infty}_{\cal
F}({\cal E}), L^{1}_{\cal F}({\cal E}))-$lower semicontinuous
conditional convex risk measure and an
$L^{p}_{\cal F}({\cal E})-$type of ${\cal T}_{\epsilon,\lambda}-$continuous conditional convex risk measure
($1\leq p<+\infty$), respectively.
 }

\vspace{1mm} \no{\footnotesize{\bf Keywords:\hspace{2mm}random normed module, random inner product module,
random locally convex module, random conjugate space, $L^{0}-$convex analysis, conditional risk measures

}}

\no{\footnotesize{\bf MSC(2000):\hspace{2mm}46A22, 46A25, 46H25, 47H40, 52A41, 91B16, 91B30, 91B70.}
 \vspace{2mm}
\baselineskip 15pt
\renewcommand{\baselinestretch}{1.22}
\parindent=10.8pt  
\rm\normalsize\rm

\section{Introduction} In the last ten years random metric
theory and its applications have undergone a systematic and deep
development, in particular random metric theory recently has been a
proper mathematical tool for the study of conditional risk measures
for unbounded financial positions. The purpose of this section is to
give a brief historic retrospect to the respective courses of the
development of random metric theory, risk measures and conditional
risk measures in order to make it easier for the reader to see how
random metric theory and the theory of conditional risk measures
come together.

\subsection{The central framework in random metric theory}

Random metric theory originated from the theory of probabilistic
metric spaces \cite{SchweizerSklar,Survey1,Survey2,Guotx-Relations}.
The central framework in random metric theory was formed in the
course of the development of random metric theory in the direction
of functional
analysis\cite{Guotx-master,Guotx-phd,You-Guo,Guotx-extension,Guo-You,Guotx-radon,Guotx-modulehom,Guotx-ref,Guo-Younote,Guotx-somebasic}.
A crucial step in the formative process was taken in
\cite{Guotx-somebasic}, where the respective new versions of the
notions of a random metric space and random normed space originally
introduced in \cite{SchweizerSklar} were presented. According to the
new versions the random norm of a vector in a random normed space
(resp., the random distance between any two points in a random
metric space) is the equivalence class of a nonnegative random
variable rather than a nonnegative random variable as defined in
\cite{SchweizerSklar}.  Since a random normed space is often endowed
with a natural topology, called the $(\epsilon,\lambda)$-topology,
it is not a locally convex linear topological space in general and the
theory of traditional conjugate spaces universally fails to serve the study of random normed spaces, for example, our recent
result in \cite{Guo-Zeng-exist} shows that for a special class of
random normed spaces---random normed modules with base
$(\Omega,{\cal F},P)$ (a probability space) they admit sufficiently
many nontrivial continuous linear functionals iff ${\cal F}$ is
generated by at most countably many $P$-atoms, so the development of
random normed spaces needs a new kind of conjugate space theory.
Motivated by the study of random linear operators in random
functional analysis and based on the earlier work in
\cite{Guotx-master,Guotx-phd,Guotx-extension,Guotx-radon}, Guo
presented in \cite{Guotx-somebasic} the definitive notion of an
almost surely bounded random linear functional and proved the
Hahn-Banach theorem for such random linear functionals, which led us
directly to the idea of random conjugate spaces. However, the
structure of random normed spaces is too weak to guarantee that an
almost surely bounded random linear functional defined on a random
normed space possesses nice properties, so that the deep development
of the theory of random conjugate spaces encounters a serious obstacle.
We found in \cite{Guotx-extension} that in order to ensure an almost
surely bounded random linear functional on a random normed space to
possess pleasant properties the random normed space has to possess a
kind of module structure, which motivated us to present the notion
of a random normed module in \cite{Guotx-extension}. Based on the
new version given in \cite{Guotx-somebasic} of a random normed
space, Guo further presented in \cite{Guotx-somebasic} the
elaborated versions of the notions of a random normed module and
random inner product module originally introduced in
\cite{Guotx-extension,Guo-You} respectively, which also leads to the
definitive notion of the random conjugate space of a random normed
space, namely, the random conjugate space of a random normed space
is exactly the random normed module consisting of all almost surely
bounded random linear functionals defined on the random normed
space. The results in \cite{Guotx-extension} show that only the
theory of random conjugate spaces for random normed modules can be
deeply developed, consequently, the center of our work has been
placed at the topics closely related to the theory of random
conjugate spaces of random normed modules since 1995. Subsequently,
we established the representation theorems of random conjugate
spaces in \cite{Guo-You,Guotx-radon} and studied module
homomorphisms on random normed modules in \cite{Guotx-modulehom}.
Motivated by the work on the representation of the dual of
Lebesgue-Bochner function spaces {\cite{Diest-Uhl}, we established
in \cite{Guotx-ref,Guotx-represent} the precise connection between
the random conjugate space $S^{\ast}$ of a random normed module $S$
and the classical conjugate space $(L^{p}(S))^{\prime}$ of the
abstract normed space $L^{p}(S)$ generated from $S$, namely
$$(L^{p}(S))^{\prime}\cong L^{q}(S^{\ast})~~~~ \bigg(1\leq
p<+\infty\quad \mbox{and}\quad \frac{1}{p}+\frac{1}{q}=1\bigg).$$
Making use of this connection, we established various
characterizations for a complete random normed module to be random
reflexive \cite{Guotx-ref,Guo-Li}, and a basic strict separation
theorem in random locally convex modules \cite{Guo-Xiao-Chen}. The
notion of a random locally convex module was introduced in
\cite{Survey2} in order to provide a proper framework for the
further development of the theory of random conjugate spaces of
random normed modules, subsequently random
$\textmd{w}^{\ast}$-topology and random weak topology were
thoroughly studied in \cite{Guotx-BBKS}, and random duality was also
developed in \cite{Guo-Chen}.

Now, random normed modules, random inner product modules and random
locally convex modules have become the central framework supporting
random metric theory, and the theory of random conjugate spaces has
been a powerful tool for the development of the central framework.
We can now say that random metric theory is being developed as
functional analysis founded on the central framework. In the course
of development, the theory of random normed modules together with
their random conjugate spaces has found many successful applications
in solving the best approximation problem and the dual
representation problem in Lebesgue-Bochner function spaces
\cite{You-Guo,Guo-Younote,Guotx-represent}, in geometry of Banach
spaces \cite{Guotx-radon,Guo-Zeng-Convexity}, and in solving various
measurability problems
\cite{Guotx-severalapplications,Guotx-skorohod}.

\subsection{Risk measures and classical convex analysis}

Risk measures were introduced in order to quantify the riskiness of
financial positions and to provide a criterion to determine whether
the risk was acceptable or not. Since Artzner et al. and Delbaen
presented and studied coherent risk measures in their seminal papers
\cite{Artzner-Delbaen-Eber,Delbaen} for the model space $L^{\infty}$
(namely the Banach space of essentially bounded random variables,
which is used to model the essentially bounded financial positions),
the theory of risk measures has obtained a quite extensive
development. In 2002, convex risk measures broader than coherent
risk measures were presented and studied by F\"ollmer and Schied in
\cite{Follmer-Schied,Follmer-Schied-robust,Follmer-Schied-Stocha}
and also independently by Frittelli and Rosazza Gianin in
\cite{Frittelli-Rosazza,Frittelli-Rosazza-Dynamic}. Since the model
space is too narrow to include the important risk models such as
normally or log-normally distributed random variables, there is a
growing mathematical finance literature dealing with convex risk
measures beyond $L^{\infty}$, see e.g.,
\cite{Delbaen-coherent,Biagini-Frittelli,Rusz-Shapiro,Kaina-Ruschen,Kratschmer,Cheridito-Li,Filipovic-Svindland}. Since convex risk
measures are extended real-valued convex functions defined on
locally convex spaces such as $L^{p}$ $(1\leq p\leq+\infty)$,
classical convex analysis\cite{Ekeland-Temam,Rockafellar} turns out
to be a powerful tool for the analysis of convex risk measures, cf.
\cite{Frittelli-Rosazza-Dynamic}.

\subsection{Conditional risk measures and random metric theory}

Conditional risk measures were introduced in order to quantify the
risk associated with financial positions when the additional
information was available. Various interpretations of the additional
information in \cite{Detlefsen-Scandolo,Bion-Nadal} show that the
theory of conditional risk measures open a way to the analysis of
the consequences of asymmetric information for risk measurement.

To briefly introduce them, let $(\Omega,{\cal E},P)$ be a
probability space, $L^{p}({\cal E})$ the Banach spaces of
equivalence classes of real-valued $p$-integrable or essentially
bounded (according to $1\leq p<+\infty$ or $p=+\infty$) ${\cal
E}$-measurable random variables on $\Omega$, $\bar{L}^{0}({\cal E})$
(or, $L^{0}({\cal E})$) the set of equivalence classes of ${\cal
E}$-measurable extended real-valued (real-valued) random variables
on $\Omega$, and ${\cal F}$ a sub $\sigma$-algebra of ${\cal E}$,
which denotes the additional information.

The first definition of a conditional risk measure, here we call it
a conditional risk measure of $L^{\infty}$-type, was introduced by
Detlefsen and Scandolo in \cite{Detlefsen-Scandolo} and
independently by Bion-Nadal in \cite{Bion-Nadal} as follows:

\th{Definition 1.3.1 {\rm\cite{Detlefsen-Scandolo,Bion-Nadal}}.}
{\it A function $f:L^{\infty}({\cal E})\rightarrow L^{\infty}({\cal
F})$ is called$:$

$(1)$ $L^{0}({\cal F})$-convex if $f(\xi x+(1-\xi)y)\leq\xi
f(x)+(1-\xi)f(y),\forall x,y\in L^{\infty}({\cal E})$ and $\xi\in
L^{0}_+({\cal F})$ such that $0\leq\xi\leq 1;$

$(2)$ monotone if $f(x)\leq f(y)$ for all $x$ and $y$ in
$L^{\infty}({\cal E})$ such that $x\geq y;$

$(3)$ cash invariant if $f(x+y)=f(x)-y$ for all $x$ in
$L^{\infty}({\cal E})$ and $y$ in $L^{\infty}({\cal F})$.

\noindent Furthermore, $f$ is called a conditional convex risk
measure of $L^{\infty}$-type if it is $L^{0}({\cal F})$-convex,
monotone and cash invariant.} \vspace{1mm}

Let ${\cal P}$ be the set of all the probability measures $Q$ on
${\cal E}$ such that $Q$ is absolutely continuous with respect to
$P$ and ${\cal P}_{\cal F}=\{Q\in {\cal P}~|~Q=P$ on ${\cal F}\}$

Given a conditional convex risk measure $f$, $\alpha:{\cal P}_{\cal
F}\rightarrow\bar{L}^{0}({\cal F})$ is defined by
$\alpha(Q)=\vee\{E_Q(-x~|~{\cal F})-f(x)~|~x\in L^{\infty}({\cal
E})\}$ for any $Q$ in ${\cal P}_{\cal F}$, called the random penalty
function of $f$, where $E_Q(\cdot~|~{\cal F})$ denotes the
conditional expectation given the $\sigma$-algebra ${\cal F}$ under
the probability $Q$. The following dual representation proposition
was proved by Detlefsen  and Scandolo in \cite{Detlefsen-Scandolo}
, see \cite{Bion-Nadal} for dual representation under a stronger assumption that $f$ is continuous from below and \cite{Follmer-Penner} for other possible forms of dual representation.

\th{Proposition 1.3.1
{\rm\cite{Detlefsen-Scandolo}}.} {\it The
following three statements are equivalent to each other$:$

$(1)$ $f(x)=\vee\{E_Q(-x~|~{\cal F})-\alpha(Q)~|~Q\in {\cal P}_{\cal
F}\},\forall x\in L^{\infty}({\cal E});$

$(2)$ $f$ is continuous from above, namely $f(x_n)\nearrow f(x)$
whenever $x_n\searrow x;$

$(3)$ $f$ has the ``Fatou property'': For any bounded sequence
$\{x_n,n\in N\}$ which converges $P$-a.s. to some $x$,
$f(x)\leq\liminf_{n\uparrow\infty}f(x_n)$.}

From the essence of the proof of Proposition 1.3.1 given in
\cite{Detlefsen-Scandolo}, one can see
that classical convex analysis may still treat the conditional
convex risk measure of $L^{\infty}$-type. Based on this kind of conditional convex risk measure, the corresponding dynamic risk measures of $L^{\infty}-$type were developed in \cite{Detlefsen-Scandolo,Follmer-Penner}, see \cite{Cheridito-Delbaen-Kupper,Cheridito-Delbaen-Eber} for $L^{\infty}-$type of dynamic risk measures of bounded stochastic processes.

Just as stated above in Section 1.2, $L^{\infty}({\cal E})$ as the
model space for conditional risk measures is too narrow. Recently,
motivated by the study of dynamic risk measures
\cite{PengS,RosazzaGianin} and conditional entropic risk measures,
Filipovi\'c, Kupper and Vogelpoth studied the following convex
conditional risk measures of $L^{p}_{\cal F}({\cal E})$-type and
$L^{p}$-type in \cite{FKV,KV,FKV-Approaches}.

Let us first introduce a conditional risk measure of $L^{p}$-type.

\th{Definition 1.3.2 {\rm\cite{FKV-Approaches}}.} {\it Let $1\leq
r\leq p<\infty$, then a function $f:L^{p}({\cal E})\rightarrow
L^{r}({\cal F})$ is called

$(1)$ $L^{0}({\cal F})$-convex if $f(\xi x+(1-\xi)y)\leq\xi
f(x)+(1-\xi)f(y),\forall x,y\in L^{p}({\cal E})$ and $\xi\in
L^{0}_+({\cal F})$ such that $0\leq\xi\leq 1;$

$(2)$ monotone if $f(x)\leq f(y)$ for all $x$ and $y$ in
$L^{p}({\cal E})$ such that $x\geq y;$

$(3)$ cash invariant if $f(x+y)=f(x)-y$ for all $x$ in $L^{p}({\cal
E})$ and $y$ in $L^{p}({\cal F});$

$(4)$ local if $\tilde{I}_A f(x)=\tilde{I}_A f(\tilde{I}_A x)$ for
all $A\in {\cal F}$ and $x\in L^{p}({\cal E})$.

\noindent Furthermore, $f$ is called a convex conditional risk
measure of $L^{p}$-type if it is convex $($namely convex in the usual
sense$)$, monotone and cash invariant.} \vspace{1mm}

Given a function $f:L^{p}({\cal E})\rightarrow L^{r}({\cal F})$, the
function $f^{\ast}:B(L^{p}({\cal E}),L^{r}({\cal F}))\rightarrow
\bar{L}^{0}({\cal F})$ is defined by
$f^{\ast}(u)=\vee\{u(x)-f(x)~|~x\in L^{p}({\cal E})\}$ for all $u\in
B(L^{p}({\cal E}),L^{r}({\cal F}))$, where $B(L^{p}({\cal
E}),L^{r}({\cal F}))$ denotes the Banach space of continuous
linear operators from $L^{p}({\cal E})$ to $L^{r}({\cal F})$,
further let $dom(f^{\ast})=\{u\in B(L^{p}({\cal E}),L^{r}({\cal
F}))~|~f^{\ast}(u)\in L^{r}({\cal F})\}$. Finally, $u\in
B(L^{p}({\cal E}),L^{r}({\cal F}))$ is called a subgradient of $f$
at $x_0\in L^{p}({\cal E})$ if $u(x-x_0)\leq f(x)-f(x_0),\forall
x\in L^{p}({\cal E})$, the set of subgradients of $f$ at $x_0$
is denoted by $\partial f(x_0)$.

Zowe proved the following in \cite{Zowe}:

\th{Proposition 1.3.2.} {\it Let $f$ be a convex function from
$L^{p}({\cal E})$ to $L^{r}({\cal F})$ and continuous at $x_0$. Then
$\partial f(x_0)\neq\emptyset$ and
$f(x_0)=\vee\{u(x_0)-f^{\ast}(u)~|~u\in
dom(f^{\ast})\}$.}\vspace{1mm}

In fact, we can prove that a convex conditional risk measure of
$L^{p}$-type is $L^{0}({\cal F})$-convex if it is continuous.
Recently, Filipovi\'c et al. proved the following in
\cite{FKV-Approaches}:

\th{Proposition 1.3.3.} {\it Let $f$ be a continuous convex
conditional risk measure from $L^{p}({\cal E})$ to $L^{r}({\cal
F})$. Then $\partial f(x)\neq\emptyset$ for all $x$ in $L^{p}({\cal
E})$ and $f(x)=\vee\{E(xy~|~{\cal F})-f^{\ast}(E(\cdot y~|~{\cal
F}))~|~y\in L^{q}({\cal E}), y\leq 0, E(y~|~{\cal F})=-1$ and
$E(|y|^{q}~|~{\cal F})\in L^{\frac{r(p-1)}{p-r}}({\cal F})\}$, where
$q$ is the H\"older conjugate number of $p$,
$\frac{r(p-1)}{p-r}=\infty$ when $p=r$ and $E(\cdot y~|~{\cal F}):
L^{p}({\cal E})\rightarrow L^{r}({\cal F})$ is defined by $E(\cdot
y~|~{\cal F})(x)=E(xy~|~{\cal F}),\forall x\in L^{p}({\cal
E})$.}\vspace{1mm}

From the essence of the proofs of Propositions 1.3.2 and 1.3.3 given in
\cite{FKV-Approaches},  the vector-valued convex analysis and a bit of basic linear operator theory can still treat a convex conditional risk measure of $L^{p}-$type.

Let $1\leq p\leq+\infty$ and $L^{p}_{{\cal F}}({\cal E})=$ the
$L^{0}({\cal F})$-module generated by $L^{p}({\cal E})$, namely
$L^{p}_{{\cal F}}({\cal E})=L^{0}({\cal F})\cdot L^{p}({\cal
E}):=\{\xi x:\xi\in L^{0}({\cal F})$ and $x\in L^{p}({\cal E})\}$,
which can be made a random normed module in a natural way. Then the
following conditional risk measure of $L^{p}_{\cal F}({\cal
E})$-type was studied in \cite{FKV,KV} and eventually presented in
\cite{FKV-Approaches}:

\th{Definition 1.3.3 {\rm\cite{FKV-Approaches}}.} {\it Let $1\leq
p\leq+\infty$. A function $f:L^{p}_{\cal F}({\cal E})\rightarrow
\bar{L}^{0}({\cal F})$ is called:

$(1)$ monotone if $f(x)\leq f(y)$ for all $x,y\in L^{p}_{\cal
F}({\cal E})$ such that $x\geq y;$

$(2)$ subcash invariant if $f(x+y)\geq f(x)-y$ for all $x\in
L^{p}_{\cal F}({\cal E})$ and $y\in L^{0}_+({\cal F});$

$(3)$ cash invariant if $f(x+y)=f(x)-y$ for all $x\in L^{p}_{\cal
F}({\cal E})$ and $y\in L^{0}({\cal F});$

\noindent Further, an $L^{0}({\cal F})$-convex, monotone and cash
invariant function from $L^{p}_{\cal F}({\cal E})$ to
$\bar{L}^{0}({\cal F})$ is called a conditional convex risk measure
of $L^{p}_{{\cal F}}({\cal E})$-type.}\vspace{1mm}

A typical example motivating Filipovi\'c, Kupper and Vogelpoth to
present and study a conditional convex risk measure  of
$L^{p}_{{\cal F}}({\cal E})$-type is the following conditional
entropic risk measure $\rho_{\gamma}(\cdot):\bar{L}^{0}({\cal
E})\rightarrow \bar{L}^{0}({\cal F})$ defined by
$\rho_{\gamma}(x)=\frac{1}{\gamma}\log E(e^{-\gamma x}~|~{\cal
F}),\forall x\in \bar{L}^{0}({\cal E})$, where $\gamma>0$ is a risk
aversion.

When $\rho_{\gamma}$ is restricted to $L^{\infty}({\cal E})$,
$\rho_{\gamma}$ is a conditional convex risk measure of
$L^{\infty}$-type continuous from above, then the dual
representation of $\rho_{\gamma}$ can be treated as in
\cite{Detlefsen-Scandolo,Bion-Nadal}.

When $\rho_{\gamma}$ is restricted to $L^{p}({\cal E})$ $(1\leq
p<+\infty)$,  $\rho_{\gamma}$ is an $L^{0}({\cal F})$-convex proper
lower semicontinuous function from $L^{p}({\cal E})$ to
$\bar{L}^{0}({\cal F})$, the methods used to establish the dual
representation results of conditional convex risk measures of
$L^{\infty}$-type and $L^{p}$-type no longer apply to
$\rho_{\gamma}$.

Motivated by the idea of hedging random future payments in a
multiperiod setting and also led by giving a more pleasant dual
representation result for $\rho_{\gamma}$ than that as given in
\cite{Detlefsen-Scandolo,Bion-Nadal}, Filipovi\'c et al. presented
in \cite{FKV} the idea of randomizing the initial data, for example,
randomizing the initial data $L^{p}({\cal E})$ into the random
normed module $L^{p}_{{\cal F}}({\cal E})$. Further, $\rho_{\gamma}$
is regarded as a mapping from $L^{p}_{{\cal F}}({\cal E})$ to
$\bar{L}^{0}({\cal F})$, then it is a conditional convex risk
measure of $L^{p}_{{\cal F}}({\cal E})$-type and is also lower
semicontinuous in the sense of \cite{FKV}. To establish the dual
representation result for this kind of conditional convex risk
measures such as $\rho_{\gamma}$ on $L^{p}_{{\cal F}}({\cal E})$,
Filipovi\'c, Kupper and Vogelpoth attempted to carry out a
spectacular generalization of the classical Fenchel-Moreau type dual
representation theorem from locally convex spaces to random locally
convex modules by substituting random locally convex modules for
classical locally convex spaces and random conjugate spaces for
classical conjugate spaces. Further, to establish the continuity and
subdifferentiability theorems for proper lower semicontinuous
$L^{0}$-convex functions on random locally convex modules, they also
introduced in \cite{FKV} the locally $L^{0}$-convex topology much
stronger than the $(\epsilon,\lambda)$-topology for random locally
convex modules. Besides, they also established in \cite{FKV} two
Hyperplane separation theorems in order to establish the
subdifferentiability and Fenchel-Moreau type dual representation
theorems for $L^{0}$-convex functions on locally $L^{0}$-convex
modules (a locally $L^{0}$-convex module amounts to a random locally
convex module endowed with the locally $L^{0}$-convex topology).

In \cite{Guotx-Relations}, Guo simultaneously considered the two
kinds of topologies---the $(\epsilon,\lambda)$-topology and locally
$L^{0}$-convex topology for a random locally convex module in order
to give the relations between the basic results currently available
derived from the two kinds of topologies. Consequently, it was
proved in \cite{Guotx-Relations} that our basic strict separation
theorem earlier established in
\cite{Guo-Xiao-Chen,Guo-Xiao,Guotx-proceeding} implies the
hyperplane separation theorem in \cite{FKV} which was used to
establish the generalized Fenchel-Moreau type dual representation
theorem in \cite{FKV}, that the two kinds of random conjugate spaces
derived from the two kinds of topologies coincide for most of random
locally convex modules, and in particular that a random locally convex
module has the same completeness under the two kinds of topologies
if the module has the countable concatenation property as defined in
\cite{Guotx-Relations}. The results in \cite{Guotx-Relations}
further show that most of the previously established principal
results of random conjugate spaces of random normed modules under
the $(\epsilon,\lambda)$-topology are still valid under the locally
$L^{0}$-convex topology, which considerably enriches financial
applications of random normed modules.

It was already pointed out in \cite{Guotx-Relations} that the new
countable concatenation property introduced in
\cite{Guotx-Relations} will play an essential role in the study of
locally $L^{0}$-convex modules. Our recent work\cite{Guo-Zhao-Zeng}
has improved the results obtained in \cite{FKV} for
separation and duality in locally $L^{0}$-convex modules. Further,
we developed the theory of random duality with respect to the
locally $L^{0}$-convex topology \cite{Guo-Zhao-Zeng}, which leads us
directly to the notion of an $L^{0}$-pre-barreled module that is
weaker than the notion of an $L^{0}$-barreled module originally
introduced in \cite{FKV}. What is important is that under the weaker
definition we can establish the characterization for a locally
$L^{0}$-convex module to be $L^{0}$-pre-barreled, which shows that a
complete random normed module with the countable concatenation
property is $L^{0}$-pre-barreled under the locally $L^{0}$-convex
topology, in particular, $L^{p}_{{\cal F}}({\cal E})$ is
$L^{0}$-pre-barreled. Therefore this weaker notion is more suitable
for the study of the continuity and subdifferentiability of a proper
lower semicontinuous conditional convex risk measure of
$L^{p}_{{\cal F}}({\cal E})$-type. In particular, we can also
establish the $(\epsilon,\lambda)$-topological version of
Fenchel-Moreau dual representation theorem for proper lower
semicontinuous $L^{0}$-convex functions \cite{Guo-Zhao-Zeng}, which
not only contains the locally $L^{0}$-convex topological version of
Fenchel-Moreau dual representation theorem established in \cite{FKV}
as a special case but also seems more natural than the latter. The
final part of this paper shows that a conditional convex risk
measure of either $L^{\infty}$-type or $L^{p}$-type can always be
extended to one of $L^{p}_{{\cal F}}({\cal E})$-type and that the
representation propositions obtained under Definition 1.3.3 have
included Propositions 1.3.1 and 1.3.3 as special cases, which
together with the
work\cite{Guotx-Relations,FKV,KV,FKV-Approaches,Guo-Zhao-Zeng} has
paved the way for unifiedly developing the theory of conditional
risk measures, so that the theory of random
locally convex modules, in particular random normed modules together
with their random conjugate spaces is a mathematical tool tailored
to the theory of conditional risk measures. We believe that the
further development of dynamic risk measures will involve more of
random metric theory.

The purpose of this paper is to give a selective survey on the
recent progress in random metric theory and its
applications to conditional risk measures.

This paper includes eight sections. The first is the very
introduction, and the remainder of this paper is organized as
follows. Section 2 gives the central framework in random metric
theory, topological structures, important examples, the notions of a
random conjugate space and the Hahn-Banach theorems for random
linear functionals; Section 3 gives several important representation
theorems for random conjugate spaces; Section 4 gives
characterizations for a complete random normed module to be random
reflexive; Section 5 gives hyperplane separation theorems currently
available in random locally convex modules; Section 6 gives the
theory of random duality with respect to the locally $L^{0}$-convex
topology and in particular a characterization for a locally
$L^{0}$-convex module to be $L^{0}$-pre-barreled; Section 7 gives
some basic results on $L^{0}$-convex analysis together with some
applications to conditional risk measures; Finally, Section 8 is
devoted to extensions of conditional risk measures.

Throughout this paper, $(\Omega,{\cal F},P)$ denotes a given
probability space, $K$ the scalar field $R$ of real numbers or $C$
of complex numbers and $L^{0}({\cal F},K)$ the algebra over $K$ of
equivalence classes of $K$-valued ${\cal F}$-measurable random
variables on $\Omega$.

Proposition 1.3.4 below is the well known theorem on the existence of an
essential supremum for a set of random variables. Let $\bar{\cal
L}^{0}({\cal F},R)$ be the set of extended real-valued ${\cal
F}$-measurable random variables on $(\Omega,{\cal F},P)$ and $H$ a
subset of $\bar{\cal L}^{0}({\cal F},R)$. $\xi\in\bar{\cal
L}^{0}({\cal F},R)$ is called an essential upper bound for $H$ if
$\eta(\omega)\leq\xi(\omega)$ $P-a.s.$ (namely $P$-almost surely)
for each $\eta\in H$; furthermore if for each essential upper bound
$\eta^{\prime}$ for $H$ it always holds that
$\eta(\omega)\leq\eta^{\prime}(\omega)$ $P-a.s.$, then the essential
upper bound $\eta$ is called an essential supremum for $H$, denoted
by ${\rm esssup}(H)$. Obviously, ${\rm esssup}(H)$ is unique
$P-a.s.$ Similarly, one has the notion of an essential infimum.

\th{Proposition 1.3.4 {\rm\cite{He-Wang-Yan}}.} {\it Every subset
$H$ of $\bar{\cal L}^{0}({\cal F},R)$ has an essential supremum and
an essential infimum, and there exist countable subsets $\{a_n,n\in
N\}$ and $\{b_n,n\in N\}$ of $H$ such that ${\rm esssup}(H)={\rm
esssup}(\{a_n,n\in N\})$ and ${\rm essinf}(H)={\rm
essinf}(\{b_n,n\in N\})$, where $N$ stands for the set of positive
integers. Furthermore, if $H$ is directed upwards $($downwards$)$
then $\{a_n,n\in N\}$ can be chosen as nondecreasing $($resp.
$\{b_n,n\in N\}$ can be chosen as nonincreasing$)$.}\vspace{1mm}

Proposition 1.3.5 below is another formulation of Proposition 1.3.4
in terms of equivalence classes, which is frequently used in random
metric theory because in random metric theory we often need to
distinguish random variables from their equivalence classes.

\th{Proposition 1.3.5 {\rm \cite{Dunford-Schwartz}}.} {\it Let
$\bar{L}^{0}({\cal F},R)$ be the set of equivalence classes of
elements in $\bar{\cal L}^{0}({\cal F},R)$, partially ordered via
$\xi\leq\eta$ iff $\xi^{0}(\omega)\leq\eta^{0}(\omega)$ $P-a.s.$,
where $\xi^{0}$ and $\eta^{0}$ are arbitrarily chosen
representatives of $\xi$ and $\eta$ in $\bar{L}^{0}({\cal F},R)$,
respectively. Then every subset $H$ of $\bar{L}^{0}({\cal F},R)$ has
a supremum, denoted by $\vee H$, and an infimum, denoted by $\wedge
H$. Furthermore, there exists a countable subset $\{a_n,n\in N\}$
$(\{b_n,n\in N\})$ of $H$ such that $\vee H=\vee_{n\geq 1}a_n$
$($resp. $\wedge H=\wedge_{n\geq 1}b_n$$)$, and $\{a_n,n\in N\}$
$($resp. $\{b_n,n\in N\}$$)$ can be chosen as nondecreasing
$($resp.,nonincreasing$)$ if $H$ is directed upwards
$($downwards$)$. Finally, ${L}^{0}({\cal F},R)$ as a sublattice of
$\bar{L}^{0}({\cal F},R)$ is complete in the sense that every subset
having an upper bound has a supremum.}\vspace{1mm}

Besides, let $\xi$ and $\eta$ be two elements in $\bar{L}^{0}({\cal
F},R)$. Then $\xi<\eta$ is understood as usual, namely $\xi\leq\eta$
and $\xi\neq\eta$. For an ${\cal F}$-measurable $A$,  $\xi<\eta$ on
$A$ is understood as $\xi^{0}(\omega)<\eta^{0}(\omega)$ $P-a.s.$ on
$A$, where $\xi^{0}$ and $\eta^{0}$ are arbitrarily chosen
representatives of $\xi$ and $\eta$, respectively.

Specially, $$L^{0}_{+}({\cal F})=\{\xi\in{L}^{0}({\cal F},R):\xi\geq
0\}$$ and $$L^{0}_{++}({\cal F})=\{\xi\in L^{0}_{+}({\cal F}):\xi>0\
\mbox{on}\ \Omega\}.$$

\section{Foundations}\vspace{-1mm}
\subsection{The central framework in random metric theory}

Basic notions in this subsection are essentially adopted from
\cite{Guotx-somebasic,Survey2}, but the changes of some notation and
convention from the theory of probabilistic metric spaces have been
made as in \cite{Guotx-Relations} in order to make our work better
known to the scholars working in other fields.

\th{Definition 2.1.1 {\rm \cite{Guotx-somebasic}}.} {\it An ordered
pair $(S,\|\cdot\|)$ is called a random normed space $($briefly, an
$RN$ space$)$ over $K$ with base $(\Omega,{\cal F},P)$ if $S$ is a
linear space over $K$ and $\|\cdot\|$ is a mapping from $S$ to
$L^{0}_{+}({\cal F})$ such that the following axioms are
satisfied$:$

$(RN$$-1)$ $\|\alpha x\|=|\alpha|\|x\|,\forall\alpha\in K$ and $x\in
S;$

$(RN$$-2)$ $\|x+y\|\leq\|x\|+\|y\|,\forall x,y\in S;$

$(RN$$-3)$ $\|x\|=0$ implies $x=\theta$ (the null vector in $S$),

\noindent where $\|x\|$ is called the random norm of the vector $x$.
Besides, a mapping $\|\cdot\|$ from $S$ to $L^{0}_{+}({\cal F})$ is
called a random seminorm on $S$ if it only satisfies $(RN$-1$)$ and
$(RN$-2$)$.

In addition, if $S$ is a left module over the algebra ${L}^{0}({\cal
F},K)$ and $\|\cdot\|$ also satisfies the following$:$

$(RNM$-1$)$ $\|\xi x\|=|\xi|\|x\|,\forall\xi\in {L}^{0}({\cal F},K)$
and $x\in S$.

\noindent Then, such an $RN$ space $(S,\|\cdot\|)$ is called a random
normed module $($briefly, an $RN$ module$)$ over $K$ with base
$(\Omega,{\cal F},P)$, such a random norm $\|\cdot\|$ is called an
${L}^{0}$-norm. Besides, a mapping $\|\cdot\|$ from $S$ to
$L^{0}_{+}({\cal F})$ is called an ${L}^{0}$-seminorm on $S$ if it
only satisfies $(RNM$-1$)$ and $(RN$-$2)$.}

\th{Remark 2.1.1.} The notion of an $RN$ module was first presented
in \cite{Guotx-extension}, the current Definition 2.1.1 is the
elaborated version of that given in \cite{Guotx-extension}. Clearly,
($RNM$-1) implies ($RN$-1). It should also be mentioned that the
notions of $L^{0}$-seminorm and $L^{0}$-norm have been known for ten
years since the notion of an $RN$ module was given in
\cite{Guotx-somebasic}, and the two notions were frequently employed
in our previous papers \cite{Survey2,Guo-Li,Guo-Peng,Guotx-proceeding}. Recently, motivated by financial applications, the
two notions of an $L^{0}$-seminorm and $L^{0}$-norm were
rediscovered in \cite{FKV,KV} and the notion of an $L^{0}$-normed
module presented in \cite{FKV,KV} is exactly the notion of a random
normed module, see \cite{Guotx-Relations} for details. Finally, the notion of an original $RN$ space was first introduced in \cite{SchweizerSklar} and the current notion of an $RN$ space was first given in \cite{Guotx-somebasic} as a new version of the notion of an original $RN$ space. By the way, if, in the definition of an $RN$ space over $R$ with base $(\Omega,{\cal F},P)$, $L^{0}({\cal F},R)$ is replaced by a complete vector lattice $Z$ and $L^{0}_{+}({\cal F})$ by $Z^{+}:=\{z\in Z~|~z\geq 0\}$, then the space obtained in such a way is called a space normed with the elements of $Z$, which was first introduced in \cite{Kantor}. Clearly, an $RN$ space is a special space normed with the elements of a complete vector lattice, but \cite{Kantor} has never been mentioned in the literature of the theory of probabilistic metric spaces perhaps since Kantorovic\cite{Kantor} did not give the notion of an $RN$ space, in particular, since Kantorovic\cite{Kantor} did not involve any discussion of randomness.

\th{Definition 2.1.2 {\rm\cite{Guotx-somebasic}}.} {\it An ordered
pair $(S,\langle\cdot,\cdot\rangle)$ is called a random inner
product space (briefly, an $RIP$ space) over $K$ with base
$(\Omega,{\cal F},P)$ if $S$ is a linear space over $K$ and
$\langle\cdot,\cdot\rangle$ is a mapping from $S\times S$ to
$L^{0}({\cal F},K)$ such that the following axioms are satisfied$:$

{\rm(RIP-1)} $\langle x,x\rangle\in L^{0}_{+}({\cal F}),\forall x\in
S$, and $\langle x,x\rangle=0$ iff $x=\theta;$

{\rm(RIP-2)} $\langle \alpha x,y\rangle=\alpha\langle
x,y\rangle,\forall \alpha\in K$ and $x,y\in S;$

{\rm(RIP-3)} $\langle x,y\rangle=\overline{\langle
y,x\rangle},\forall x,y\in S$, where $\overline{\langle y,x\rangle}$
denotes the complex conjugate of $\langle y,x\rangle;$

{\rm(RIP-4)} $\langle x+y,z\rangle=\langle x,z\rangle+\langle
y,z\rangle,\forall x,y,z\in S$,

\noindent where $\langle x,y\rangle$ is called the random inner
product from $x$ to $y$.

In addition, if $S$ is a left module over the algebra $L^{0}({\cal
F},K)$ and the following axiom is also satisfied$:$

$(RIPM$-1$)$ $\langle \xi x,y\rangle=\xi\langle
x,y\rangle,\forall\xi\in L^{0}({\cal F},K)$ and $x,y\in S$,

\noindent then such an $RIP$ space $(S,\langle \cdot,\cdot\rangle)$
is called a random inner product module $($briefly, an $RIP$
module$)$ over $K$ with base $(\Omega,{\cal F},P)$, and $\langle
\cdot,\cdot\rangle$ is called an $L^{0}$-inner product on $S$.}

\th{Proposition 2.1.1 {\rm \cite{Guotx-somebasic}}.} {\it Let
$(S,\langle \cdot,\cdot\rangle)$ be an $RIP$ space over $K$ with
base $(\Omega,{\cal F},P)$. Define $\|\cdot\|:S\rightarrow
L^{0}_{+}({\cal F})$ by $\|x\|=\sqrt{\langle x,x\rangle},\forall
x\in S$. Then we have the following statements$:$

$(1)$ $($Cauchy-Schwartz inequality$)$ $|\langle
x,y\rangle|\leq\|x\|\|y\|,\forall x,y\in S;$

$(2)$ $(S,\|\cdot\|)$ is an $RN$ space over $K$ with base
$(\Omega,{\cal F},P)$, and is an $RN$ module if $(S,\langle
\cdot,\cdot\rangle)$ is an $RIP$ module$;$

$(3)$ A random norm $\|\cdot\|$ on a linear space $S$ is derivable
from some random inner product $\langle \cdot,\cdot\rangle$ on $S$
as above iff $\|\cdot\|$ satisfies the parallelogram law, namely
$\|x+y\|^{2}+\|x-y\|^{2}=2\|x\|^{2}+2\|y\|^{2},\forall x,y\in S$.}

\th{Definition 2.1.3 {\rm\cite{Survey2}}.} {\it An ordered pair
$(S,{\cal P})$ is called a random locally convex space over $K$ with
base $(\Omega,{\cal F},P)$ if $S$ is a linear space over $K$ and
${\cal P}$ is a family of random seminorms on $S$ such that the
following axiom is satisfied$:$

$(1)$ $\vee\{\|x\|:\|\cdot\|\in{\cal P}\}=0$ iff $x=\theta$.

In addition, if $S$ is a left module over the algebra $L^{0}({\cal
F},K)$ and each $\|\cdot\|$ in ${\cal P}$ is an $L^{0}$-seminorm,
then such a random locally convex space is called a random locally
convex module over $K$ with base $(\Omega,{\cal F},P)$.}

\th{Remark 2.1.2.}The terminologies ``random locally convex modules
and random locally convex spaces'' were first employed in
\cite{Guo-Xiao-Chen}, and they were called ``random seminormed modules
and random seminormed spaces'' in our previous papers
\cite{Guotx-modulehom,Survey2,Guo-Peng,Guo-Zhu}. When ${\cal P}$
reduces to a singleton $\{\|\cdot\|\}$, then a random locally convex
module $(S,{\cal P})$ is exactly an $RN$ module $(S,\|\cdot\|)$.
Clearly, $L^{0}({\cal F},K)$ is an $RN$ module over $K$ with base
$(\Omega,{\cal F},P)$ endowed with the $L^{0}$-norm $\|\cdot\|$
defined by $\|x\|=|x|,\forall x\in L^{0}({\cal F},K)$, where $|x|$
is the equivalence class of the composite function
$|x^{0}|:\Omega\rightarrow [0,+\infty)$ defined by
$|x^{0}|(\omega)=|x^{0}(\omega)|,\forall\omega\in\Omega$, while
$x^{0}$ is an arbitrarily chosen representative of $x$.

\subsection{Topological structures}

The $(\epsilon,\lambda)$-topology was first introduced by Schweizer
and Sklar in 1961 for probabilistic metric spaces, see
\cite{SchweizerSklar} for details. It is naturally and frequently used
in probability theory since the $(\epsilon,\lambda)$-topology for
$L^{0}({\cal F},K)$ is exactly the well known topology of
convergence in probability $P$. Definition 2.2.2 below merely
employed the idea of Schweizer and Sklar's introduction of the
$(\epsilon,\lambda)$-topology for probabilistic metric spaces.

In the sequel, for a random locally convex space $(S,{\cal P})$ with
base $(\Omega,{\cal F},P)$ and for each finite subfamily $Q$ of
${\cal P}$, $\|\cdot\|_{Q}:S\rightarrow L^{0}_{+}({\cal F})$ always
denotes the random seminorm of $S$ defined by
$\|x\|_{Q}=\vee\{\|x\|~|~\|\cdot\|\in Q\},\forall x\in S$, and
${\cal F}({\cal P})$ the set of finite subfamilies of ${\cal P}$.

Besides, we always follow the convention from random metric theory:
for each $A\in{\cal F}$, $I_A$ stands for the characteristic
function of $A$, and $\tilde{I}_A$ its equivalence class.

\th{Definition 2.2.1.} {\it Let $(S,{\cal P})$ be a random locally
convex space over $K$ with base $(\Omega,{\cal F},P)$. For each
countable subset $\{Q_n~|~n\in N\}$ of ${\cal F}({\cal P})$ and each
countable partition $\{A_n~|~n\in N\}$ of $\Omega$ to ${\cal F}$
$($namely each $A_n\in{\cal F},\bigcup_{n\geq 1}A_n=\Omega$, and
$A_n\cap A_m=\emptyset$ when $m\neq n)$, the random seminorm
$\|\cdot\|:S\rightarrow L^{0}_{+}({\cal F})$ defined by
$\|x\|=\sum_{n\geq 1}\tilde{I}_{A_n}\|x\|_{Q_n},\forall x\in S,$ is
called a countable concatenation in ${\cal P}$. Denote by ${\cal
P}_{cc}$ the set of countable concatenations in ${\cal P}$. ${\cal
P}$ is said to have the countable concatenation property if ${\cal
P}_{cc}={\cal P}$.}

\th{Remark 2.2.1.}The countable concatenation property for the
family of $L^{0}$-seminorms was first introduced in \cite{FKV}.
Readers can easily see that ${\cal P}_{cc}$ plays the same role as
the saturation of ${\cal P}$ introduced in \cite{Guo-Chen}. Clearly,
$(S,{\cal P}_{cc})$ is also a random locally convex space.

\th{Definition 2.2.2 {\rm\cite{Survey2,Guotx-modulehom}}.} {\it Let
$(S,{\cal P})$ be a random locally convex space over $K$ with base
$(\Omega,{\cal F},P)$. For any positive real numbers $\epsilon$ and
$\lambda$ such that $0<\lambda<1$, and any $Q\in{\cal F}({\cal P})$,
let $N_{\theta}(Q,\epsilon,\lambda)=\{x\in
S~|~P\{\omega\in\Omega~|~\|x\|_Q(\omega)<\epsilon\}>1-\lambda\}$,
then $\{N_{\theta}(Q,\epsilon,\lambda)~|~Q\in{\cal F}({\cal
P}),\epsilon>0,0<\lambda<1\}$ is easily verified to be a local base
at the null vector $\theta$ of some Hausdorff linear topology. The
linear topology is called the $(\epsilon,\lambda)$-topology
introduced by ${\cal P}$.}\vspace{1mm}

From now on, the $(\epsilon,\lambda)$-topology for each random
locally convex space is always denoted by ${\cal
T}_{\epsilon,\lambda}$ whenever no confusion occurs.

\th{Proposition 2.2.1 {\rm\cite{Survey2}}.} {\it Let $(S,{\cal P})$
be a random locally convex space over $K$ with base $(\Omega,{\cal
F},P)$. Then we have the following statements$:$

$(1)$ The $(\epsilon,\lambda)$-topology induced by ${\cal P}$ is the
same as the one induced by ${\cal P}_{cc};$

$(2)$ The $(\epsilon,\lambda)$-topology for $L^{0}({\cal F},K)$ is
exactly the topology of convergence in probability $P$, and
$(L^{0}({\cal F},K),{\cal T}_{\epsilon,\lambda})$ is a topological
algebra over $K;$

$(3)$ If $(S,{\cal P})$ is a random locally convex module, then
$(S,{\cal T}_{\epsilon,\lambda})$ is a topological module over the
topological algebra $L^{0}({\cal F},K);$

$(4)$ A net $\{x_{\delta},\delta\in\Gamma\}$ converges in the
$(\epsilon,\lambda)$-topology to some $x$ in $S$ iff for each
$\|\cdot\|\in{\cal P}$ $\{\|x_{\delta}-x\|,\delta\in\Gamma\}$
converges in probability $P$ to $0$.}\vspace{1mm}

The following locally $L^{0}$-convex topology is easily seen to be
much stronger than the $(\epsilon,\lambda)$-topology, and was first
introduced by Filipovi\'c et al. in \cite{FKV} for random locally
convex modules.

{\th{Definition 2.2.3 {\rm\cite{FKV}}.} {\it Let $(S,{\cal P})$ be a
random locally convex space over $K$ with base $(\Omega,{\cal
F},P)$. For any $Q\in{\cal F}({\cal P})$ and $\epsilon\in
L^{0}_{++}({\cal F})$, let $$N_{\theta}(Q,\epsilon)=\{x\in
S~|~\|x\|_Q\leq\epsilon\}.$$ A subset $G$ of $S$ is called ${\cal
T}_c$-open if for each $x\in G$ there exists some
$N_{\theta}(Q,\epsilon)$ such that $x+N_{\theta}(Q,\epsilon)\subset
G$, ${\cal T}_c$ denotes the family of ${\cal T}_c$-open subsets of
$S$. Then it is easy to see that $(S,{\cal T}_c)$ is a Hausdorff
topological group with respect to the addition on $S$. ${\cal T}_c$
is called the locally $L^{0}$-convex topology induced by ${\cal
P}$.}

\th{Remark 2.2.2.}${\cal T}_c$ as in Definition 2.2.3 is not
necessarily a linear topology. As shown in \cite{FKV}, the locally
$L^{0}$-convex topology ${\cal T}_c$ for the algebra $L^{0}({\cal
F},K)$ are only a topological ring in general, for example, the
mapping $\alpha\rightarrow\alpha\cdot 1$ may be not continuous in
general, where 1 denotes the multiplication unit element of
$L^{0}({\cal F},K)$.\vspace{1mm}

From now on, the locally $L^{0}$-convex topology for each random
locally convex space is always denoted by ${\cal T}_c$ whenever no
confusion exists.

\th{Proposition 2.2.2 {\rm \cite{FKV}}.} {\it Let $(S,{\cal P})$ be
a random locally convex module over $K$ with base $(\Omega,{\cal
F},P)$. Then $(S,{\cal T}_c)$ is a Hausdorff topological module over
the topological ring $(L^{0}({\cal F},K),{\cal T}_c)$ and
$\{N_{\theta}(Q,\epsilon)~|~Q\in{\cal F}({\cal P}),\epsilon\in
L^{0}_{++}({\cal F})\}$ is just a local base at $\theta$ of ${\cal
T}_c$.}\vspace{1mm}

Topological modules over the topological algebra $((L^{0}({\cal
F},K),{\cal T}_{\epsilon,\lambda})$ were earlier and deeply studied
in \cite{Guo-Peng}, whereas topological modules over the topological
ring $(L^{0}({\cal F},K),{\cal T}_c)$ were only recently studied in
\cite{FKV}. To introduce an important result of \cite{FKV}, we first
give Definition 2.2.4 below, which was independently introduced in
\cite{FKV} and \cite{Guo-Chen}, in particular, the notion of an
$L^{0}$-convex set almost occurred in all our previous work, see
e.g.,
\cite{Guotx-modulehom,Guotx-somebasic,Guo-Li,Guo-Xiao-Chen,Guo-Xiao}
in the name of an $M$-convex set.

\th{Definition 2.2.4 {\rm\cite{Guo-Chen,FKV}}.} {\it Let $S$ be a
left module over the algebra $L^{0}({\cal F},K)$ and $A$ a subset of
$S$. Then $A$ is called $L^{0}$-convex if $\xi x+\eta y\in A$ for
any $x$ and $y$ in $A$ and for any $\xi$ and $\eta$ in
$L^{0}_{+}({\cal F})$ such that $\xi+\eta=1$. $A$ is called
$L^{0}$-absorbent if for each $x\in S$ there exists some $\xi\in
L^{0}_{++}({\cal F})$ such that $x\in\xi A:=\{\xi a~|~a\in A\}$. $A$
is called $L^{0}$-balanced if $\xi x\in A$ for any $x\in A$ and
$\xi\in L^{0}({\cal F},K)$ such that $|\xi|\leq 1$.}

\th{Definition 2.2.5 {\rm\cite{FKV}}.} {\it A topological module
$(S,{\cal T})$ over the topological ring $(L^{0}({\cal F},K),{\cal
T}_c)$ is called a locally $L^{0}$-convex module over $K$ with base
$(\Omega,{\cal F},P)$ if there is a local base ${\cal U}_{\theta}$
at $\theta$ for ${\cal T}$ such that each element in ${\cal
U}_{\theta}$ is $L^{0}$-convex, $L^{0}$-absorbent and
$L^{0}$-balanced.}\vspace{1mm}

Clearly, for each random locally convex module $(S,{\cal P})$,
$(S,{\cal T}_c)$ is a Hausdorff locally $L^{0}$-convex module. What
is important is that Filipovi\'c et al. made use of the notion of a
gauge function to obtain the following:

\th{Proposition 2.2.3 {\rm\cite{FKV}}.} {\it A topological module
$(S,{\cal T})$ over the topological ring $(L^{0}({\cal F},K),{\cal
T}_c)$ is a locally $L^{0}$-convex module iff ${\cal T}$ can be
induced by a family of $L^{0}$-seminorms on $S$.}\vspace{1mm}

Thus the theory of Hausdorff locally $L^{0}$-convex modules amounts
to that of random locally convex modules endowed with the locally
$L^{0}$-convex topology, namely the locally $L^{0}$-convex topology
perfectly matches the family of $L^{0}$-seminorms. Besides, the
locally $L^{0}$-convex topology has one more advantage: it is strong
enough to play a crucial role in continuity and subdifferentiability
theorems for lower semicontinuous $L^{0}$-valued proper functions,
see Section 7 for details. On the other hand, it is also too strong
to make many basic theorems concerning separation and duality in
locally $L^{0}$-convex modules valid unless these modules possess a
new countable concatenation property as introduced in
\cite{Guotx-Relations} and Definition 2.2.6 below, see Sections 5
and 7 for details. Finally, the locally $L^{0}$-convex topology does
not possess some pleasant properties as in Proposition 2.2.4
below, but the $(\epsilon,\lambda)$-topology can complement these
drawbacks of the locally $L^{0}$-convex topology. Consequently, in
the long run, the two kinds of topologies should be simultaneously
considered in the future development of random locally convex
modules, just as pointed out in \cite{Guotx-compre}.

Let $(S,\|\cdot\|)$ be an $RN$ module over $K$ with base
$(\Omega,{\cal F},P)$ and $p$ an extended positive number such that
$1\leq p\leq+\infty$. Define $\|\cdot\|_p:S\rightarrow[0,+\infty]$
as follows for $x$ in $S$:
\[
\|x\|_p=\left\{
\begin{array}{ll}
\bigg(\int_{\Omega}{(\|x\|)^p}d P\bigg)^{\frac {1}{p}}, &\mbox{when
~$1\leq
p<+\infty$~,}\\
\mbox{the $P$-essential supremum of $\|x\|$}, &\mbox{$p=+\infty$}.
\end{array}
 \right.
\]

Denote $\{x\in S~|~\|x\|_p<+\infty\}$ by $L^{p}(S)$, then
$(L^{p}(S),\|\cdot\|_p)$ is a normed space over $K$, and a Banach
space if $S$ is ${\cal T}_{\epsilon,\lambda}$-complete. It is easy
to see that $L^{p}(S)$ is just the ordinary Lebesgue-Bochner
function space $L^{p}({\cal F},B)$ when $S$ is the $RN$ module
$L^{0}({\cal F},B)$ of equivalence classes of ${\cal F}$-random
variables from $(\Omega,{\cal F},P)$ to a normed space $B$ (see
Section 2.3 below).

Although one can define $L^{p}(S)$ for an $RN$ space
$(S,\|\cdot\|)$, $L^{p}(S)$ may be trivial, namely
$L^{p}(S)=\{\theta\}$. Proposition 2.2.4 below shows that this would
not occur for an $RN$ module $(S,\|\cdot\|)$ and also reflects the
importance of the $(\epsilon,\lambda)$-topology.

\th{Proposition 2.2.4
{\rm\cite{Guotx-ref,Guotx-represent,Guo-Li,Guo-Xiao-Chen}}.} {\it
$L^{p}(S)$ is ${\cal T}_{\epsilon,\lambda}$-dense in every $RN$
module $(S,\|\cdot\|)$ for each $p$ such that $1\leq
p\leq+\infty$.}\vspace{1mm}

Definition 2.2.6 below, namely the notion of the countable
concatenation property of a left module over the algebra
$L^{0}({\cal F},K)$ was first introduced in \cite{Guotx-Relations},
which is different from either of the two kinds of countable
concatenation properties introduced in \cite{FKV}. In fact, the two
kinds of countable concatenation properties in the sense of
\cite{FKV} are essentially identical, namely the countable
concatenation property of a family of $L^{0}$-seminorms as
introduced in Definition 2.2.1.

Let $S$ be a left module over the algebra $L^{0}({\cal F},K)$. For
any given countable subset $\{x_n,n\in N\}$ and any given countable
partition $\{A_n,n\in N\}$ of $\Omega$ to ${\cal F}$, the countable
formal sum $\sum_{n\geq 1}\tilde{I}_{A_n}x_n$ is not defined in
general, where we call it a countable concatenation. If  $\{x_n,n\in
N\}$ is contained in a subset $G$ of $S$, then the countable
concatenation $\sum_{n\geq 1}\tilde{I}_{A_n}x_n$ is called a
countable concatenation from $G$. For any two countable
concatenation $\sum_{n\geq 1}\tilde{I}_{A_n}x_n$ and $\sum_{n\geq
1}\tilde{I}_{B_n}y_n$, we say that they are equal if
$\tilde{I}_{A_i\cap B_j}x_i=\tilde{I}_{A_i\cap B_j}y_j$ for any $i$
and $j$ in $N$. Finally, we say that a countable concatenation
$\sum_{n\geq 1}\tilde{I}_{A_n}x_n$ is well defined if there exists
$x\in S$ such that $\tilde{I}_{A_n}x=\tilde{I}_{A_n}x_n$ for any
$n\in N$. In this paper, we always make the convention for a module
$S$ over the algebra $L^{0}({\cal F},K)$: for any two elements $x$
and $y$ in $S$, if there exists a countable partition $\{A_n,n\in
N\}$ of $\Omega$ to ${\cal F}$ such that
$\tilde{I}_{A_n}x=\tilde{I}_{A_n}y$ for each $n\in N$, then $x=y$.
It is easy to see that any random locally convex module $(S,{\cal
P})$ satisfies the above convention.

\th{Definition 2.2.6 {\rm\cite{Guotx-Relations}}.} {\it Let $S$ be a
left module over the algebra $L^{0}({\cal F},K)$ and $G$ a subset of
$S$. The set of countable concatenations from $G$, denoted by
$G_{cc}$, is called the countable concatenation hull of $G$. If
$G_{cc}=G$, then $G$ is called having the countable concatenation
property, namely $G$ is closed under the countable concatenation
operation, that is to say, for any countable concatenation
$\sum_{n\geq 1}\tilde{I}_{A_n}x_n$ from $G$, there exists $x$ in $G$
such that $\tilde{I}_{A_n}x=\tilde{I}_{A_n}x_n$, for any $n\in N$.
Specially, if $S_{cc}=S$, then $S$ is called having the countable
concatenation property.}\vspace{1mm}

Now, we can say that the countable concatenation property is
ubiquitous in random metric theory: for any ${\cal
T}_{\epsilon,\lambda}$-complete random locally convex module
$(S,{\cal P})$, $S$ has the countable concatenation property; The
random conjugate space $E_{\epsilon,\lambda}^{\ast}$ (see Section
2.4) of a random locally convex module under the
$(\epsilon,\lambda)$-topology also has the countable concatenation
property; For all random normed modules $(S,\|\cdot\|)$ occurring in
random analysis and the study of conditional risk measures (see
Section 2.3), $S$ has the countable concatenation property. It
turned out that the deep advances in the theory of $RN$ modules
together with their random conjugate
spaces\cite{Guotx-Relations,Guo-You,Guotx-radon,Guotx-ref,Guo-Li,Guo-Xiao-Chen,Guo-Chen,Guotx-randomduality}
just benefited from the countable concatenation property.

Propositions 2.2.5 and 2.2.6 below are of fundamental importance in
the theory of random locally convex modules.

\th{Proposition 2.2.5 {\rm \cite{Guotx-Relations}}.} {\it Let
$(S,{\cal P})$ be a random locally convex module and $A$ a subset
having the countable concatenation property of $S$. Then
$\bar{A}_c=\bar{A}_{\epsilon,\lambda}$, where $\bar{A}_c$ and
$\bar{A}_{\epsilon,\lambda}$ stand for the ${\cal T}_c$-closure and
${\cal T}_{\epsilon,\lambda}$-closure of $A$, respectively.
Specially, $A$ is ${\cal T}_c$-closed iff $A$ is ${\cal
T}_{\epsilon,\lambda}$-closed.}

\th{Proposition 2.2.6 {\rm \cite{Guotx-Relations}}.} {\it Let
$(S,{\cal P})$ be a random locally convex module. Then $S$ is ${\cal
T}_c$-complete if $S$ is ${\cal T}_{\epsilon,\lambda}$-complete.
Furthermore, if $S$ has the countable concatenation property, then
$S$ is ${\cal T}_c$-complete iff $S$ is ${\cal
T}_{\epsilon,\lambda}$-complete.}

\subsection{Important examples}

Let us first recall from \cite{Neveu}: Let $(B,\|\cdot\|)$ be a
normed space over $K$. Then a mapping $x^{0}:(\Omega,{\cal
F},P)\rightarrow (B,\|\cdot\|)$ is called a $B$-valued ${\cal
F}$-random variable on $\Omega$ if it is the pointwise limit of a
sequence of simple $B$-valued ${\cal F}$-measurable mapping on
$\Omega$. Further, $B^{\prime}$ denotes the classical conjugate
space of $B$, then a mapping $q^{0}:(\Omega,{\cal F},P)\rightarrow
B^{\prime}$ is called a $\textmd{w}^{\ast}$-random variable if the
composite function $\langle b,q^{0}\rangle:\Omega\rightarrow K$
defined by $\langle b,q^{0}\rangle(\omega)=\langle
b,q^{0}(\omega)\rangle:=((q^{0})(\omega))(b),\forall\omega\in\Omega$,
is a $K$-valued random variable for any given $b\in B$. For a
$\textmd{w}^{\ast}$-random variable $q^{0}$, the nonnegative
function $\|q^{0}\|:\Omega\rightarrow[0,+\infty)$ defined by
$\|q^{0}\|(\omega)=\|q^{0}(\omega)\|,\forall\omega\in\Omega$, is not
necessarily ${\cal F}$-measurable, but
$\xi_{q^{0}}:=esssup\{|\langle b,q^{0}\rangle|~|~b\in B$ and
$\|b\|\leq 1\}$ is always a nonnegative real-valued ${\cal
F}$-measurable function on $\Omega$.

\th{Example 2.3.1 {\rm\cite{Guotx-radon}}.} Denote by $L^{0}({\cal
F},B)$ the linear space of equivalence classes of $B$-valued ${\cal
F}$-random variables on $\Omega$. The module multiplication
operation $\cdot:L^{0}({\cal F},K)\times L^{0}({\cal
F},B)\rightarrow L^{0}({\cal F},B)$ is defined by $\xi x=$ the
equivalence class of $\xi^{0}x^{0}$, where $\xi^{0}$ and $x^{0}$ are
the respective arbitrarily chosen representatives of $\xi\in
L^{0}({\cal F},K)$ and $x\in L^{0}({\cal F},B)$, and
$(\xi^{0}x^{0})(\omega)=\xi^{0}(\omega)\cdot
x^{0}(\omega),\forall\omega\in\Omega$. Furthermore, the mapping
$\|\cdot\|:L^{0}({\cal F},B)\rightarrow L^{0}_+({\cal F})$ is
defined by $\|x\|=$ the equivalence class of $\|x^{0}\|,\forall x\in
L^{0}({\cal F},B)$, where $x^{0}$ is as above. Then it is easy to
see that $(L^{0}({\cal F},B),\|\cdot\|)$ is an $RN$ module over $K$
with base $(\Omega,{\cal F},P)$.

\th{Example 2.3.2 {\rm \cite{Guotx-radon,Survey1}}.}Denote by
$L^{0}({\cal F},B^{\prime},\textmd{w}^{\ast})$ the linear space of
$\textmd{w}^{\ast}$-equivalence classes of $B^{\prime}$-valued
$\textmd{w}^{\ast}$-random variables on $\Omega$. The module
multiplication operation on $L^{0}({\cal
F},B^{\prime},\textmd{w}^{\ast})$ is similarly defined as in the
case of $L^{0}({\cal F},B)$. Furthermore, the mapping
$\|\cdot\|:L^{0}({\cal F},B^{\prime},\textmd{w}^{\ast})\rightarrow
L^{0}_+({\cal F})$ is defined by $\|q\|=$ the equivalence class of
$\xi_{q^{0}}$, where $q^{0}$ is an arbitrarily chosen representative
of $q\in L^{0}({\cal F},B^{\prime},\textmd{w}^{\ast})$. Then it is
easy to see that $(L^{0}({\cal
F},B^{\prime},\textmd{w}^{\ast}),\|\cdot\|)$ is an $RN$ module over
$K$ with base $(\Omega,{\cal F},P)$.

\th{Example 2.3.3 {\rm \cite{Guotx-Relations}}.}Let $(S,\|\cdot\|)$
be an $RN$ module over $K$ with base $(\Omega,{\cal E},P)$ and
${\cal F}$ a sub $\sigma$-algebra of ${\cal E}$.
$|||\cdot|||_p:S\rightarrow \bar{L}^{0}_+({\cal F})$ is defined as
follows for $x\in S$ and $p\in[1,+\infty]$:
\[
|||x|||_p=\left\{
\begin{array}{ll}
[E(\|x\|^{p}~|~{\cal F})]^{\frac {1}{p}}, &\mbox{when ~$1\leq
p<+\infty$~,}\\
\wedge\{\xi\in \bar{L}^{0}_+({\cal F})~|~\xi\geq\|x\|\},
&\mbox{$p=+\infty$},
\end{array}
 \right.
\]
where $E(\|x\|^{p}~|~{\cal
F})=\lim_{n\rightarrow\infty}E(\|x\|^{p}\wedge n~|~{\cal F})$
denotes the extended conditional expectation. Let $L^{p}_{\cal
F}(S)=\{x\in S~|~|||x|||_p\in {L}^{0}_+({\cal F})\}$, then it is
easy to see that $(L^{p}_{\cal F}(S),|||\cdot|||_p)$ is an $RN$
module over $K$ with base $(\Omega,{\cal F},P)$.\vspace{1mm}

The method to construct Example 2.3.3 comes from \cite{FKV} where an
extremely important $RN$ module $L^{p}_{\cal F}({\cal E})$ was
constructed. Since $L^{p}_{\cal F}({\cal E})$ has been used as the
model space for conditional risk measures in \cite{FKV-Approaches},
it should be given as follows:

\th{Example 2.3.4 {\rm\cite{FKV}}.} Let $(\Omega,{\cal E},P)$ be a
probability space and ${\cal F}$ a sub $\sigma$-algebra of ${\cal
E}$. Take $S=L^{0}({\cal E},R)$ in Example 2.3.3, then $L^{p}_{\cal
F}(S)$ is exactly $L^{p}_{\cal F}({\cal E})$ as constructed in
\cite{FKV}.

\subsection{Random conjugate spaces and Hahn-Banach extension theorems for random linear functionals}

\th{Definition 2.4.1 {\rm \cite{Survey1}}.} {\it Let $S$ be a linear
space over $K$. Then a linear operator from $S$ to $L^{0}({\cal
F},K)$ is called a random linear functional on $S$. Furthermore if
$S$ is a left module over the algebra $L^{0}({\cal F},K)$, then a
module homomorphism from $S$ to $L^{0}({\cal F},K)$ is called an
$L^{0}$-linear function.}

\th{Definition 2.4.2 {\rm \cite{Guotx-Relations}}.} {\it Let $S$ be
a real linear space. A mapping $f$ from $S$ to $L^{0}({\cal F},R)$
is called a random sublinear functional on $S$ if the following are
satisfied$:$

$(1)$ $f(\alpha x)=\alpha f(x),\forall\alpha\geq 0$ and $x\in S;$

$(2)$ $f(x+y)\leq f(x)+f(y),\forall x,y\in S$.

\noindent Furthermore, if $S$ is a left module over the algebra
$L^{0}({\cal F},R)$, then a mapping $f$ from $S$ to $L^{0}({\cal
F},R)$ is called $L^{0}$-sublinear function on $S$ if it satisfies
the above $(2)$ and the following$:$

$(3)$ $f(\xi x)=\xi f(x),\forall\xi\in L^{0}_+({\cal F})$ and $x\in
S$.}

\th{Definition 2.4.3 {\rm \cite{Guotx-somebasic}}.} {\it Let
$(S,\|\cdot\|)$ be an $RN$ space over $K$ with base $(\Omega,{\cal
F},P)$. A random linear functional $f:S\rightarrow L^{0}({\cal
F},K)$ is called a.s. bounded if there exists some $\xi\in
L^{0}_+({\cal F})$ such that $|f(x)|\leq\xi\|x\|,\forall x\in S$.
Denote by $S^{\ast}$ the set of a.s. bounded random linear
functionals on $S$. The module multiplication operation
$\cdot:L^{0}({\cal F},K)\times S^{\ast}\rightarrow S^{\ast}$ is
defined by $(\xi f)(x)=\xi (f(x)),\forall\xi\in L^{0}({\cal
F},K),f\in S^{\ast}$ and $x\in S$, and the mapping
$\|\cdot\|^{\ast}:S^{\ast}\rightarrow L^{0}_+({\cal F})$ is defined
by $\|f\|^{\ast}=\wedge\{\xi\in L^{0}_+({\cal F})~|~|f(x)|\leq\xi
\|x\|,\forall x\in S\}$. It is easy to see that
$(S^{\ast},\|\cdot\|^{\ast})$ is an $RN$ module over $K$ with base
$(\Omega,{\cal F},P)$, called the random conjugate space of
$(S,\|\cdot\|)$.}\vspace{1mm}

Let $\Omega=[0,1],{\cal F}$ be the $\sigma$-algebra of Lebesgue
measurable subsets of [0,1] and $P$ be the Lebesgue measure, then it
is well known that there is no nontrivial continuous linear
functional on $(L^{0}({\cal F},K),{\cal T}_{\epsilon,\lambda})$. But
$L^{0}({\cal F},K)$ always has its random conjugate space (in fact,
$(L^{0}({\cal F},K))^{\ast}=L^{0}({\cal F},K)$, see Section 3),
further, Proposition 2.4.3 below justifies the theory of random
conjugate spaces.

Proposition 2.4.1 below was first given in \cite{Guotx-master} in
the context of random linear functionals, whose proof was merely a
copy of the Hahn-Banach theorem for real linear functionals by
noticing the order completeness of $L^{0}({\cal F},R)$. In fact,
Proposition 2.4.1 is known as a special case of more general results
in \cite{Breckner-Scheiber,Vuza}.

\th{Proposition 2.4.1.} {\it Let $S$ be a real linear space,
$M\subset S$ a subspace, $f:M\rightarrow L^{0}({\cal F},R)$ a random
linear functional and $p:S\rightarrow L^{0}({\cal F},R)$ a random
sublinear functional such that $f(x)\leq p(x),\forall x\in M$. Then
there exists a random linear functional $g:S\rightarrow L^{0}({\cal
F},R)$ such that $g$ extends $f$ and $g(x)\leq p(x),\forall x\in
S$.}\vspace{1mm}

Proposition 2.4.2 below was essentially proved in
\cite{Guotx-master} in an indirect manner, whose direct proof was
recently given in \cite{Guotx-Relations}.

\th{Proposition 2.4.2 {\rm \cite{Guotx-Relations,Guotx-master}}.}
{\it Let $S$ be a complex linear space, $M\subset S$ a subspace,
$f:M\rightarrow L^{0}({\cal F},C)$ a random linear functional and
$p:S\rightarrow L^{0}_+({\cal F})$ a random seminorm such that
$|f(x)|\leq p(x),\forall x\in M$. Then there exists a random linear
functional $g:S\rightarrow L^{0}({\cal F},C)$ such that $g$ extends
$f$ and $|g(x)|\leq p(x),\forall x\in S$.}

\th{Proposition 2.4.3 {\rm\cite{Guotx-master}}.} {\it Let
$(S,\|\cdot\|)$ be an $RN$ space over $K$ with base $(\Omega,{\cal
F},P)$, $M\subset S$ a subspace and $f:M\rightarrow L^{0}({\cal
F},K)$ an a.s. bounded random linear functional. Then there exists
$g\in S^{\ast}$ such that $g$ extends $f$ and
$\|f\|^{\ast}=\|g\|^{\ast}$.}\vspace{1mm}

Armed with the notion of a random conjugate space (namely Definition
2.4.3) and the Hahn-Banach theorem (namely Proposition 2.4.3), we
knew that lots of basic results in classical functional analysis
could be translated to $RN$ spaces. To prove the ${\cal
T}_{\epsilon,\lambda}$-completeness of $S^{\ast}$ for any $RN$ space
$(S,\|\cdot\|)$, we first translated the theory of bounded linear
operators, which led to the subject of
\cite{Guotx-phd,Guotx-extension,Guotx-modulehom}.

For the sake of convenience, in the sequel we always use $\|\cdot\|$
for the random norm on any $RN$ space if no confusion produces.

\th{Definition 2.4.4 {\rm \cite{Guotx-phd,Guotx-modulehom}}.} {\it
Let $E$ and $F$ be any two $RN$ spaces over $K$ with base
$(\Omega,{\cal F},P)$. A linear operator $T:E\rightarrow F$ is
called a.s. bounded if there exists some $\xi\in L^{0}_{+}({\cal
F})$ such that $\|Tx\|\leq\xi\|x\|,\forall x\in E$. Denote by
$B(E,F)$ the linear space of a.s. bounded linear operators from $E$
to $F$, define $\|\cdot\|:B(E,F)\rightarrow L^{0}_{+}({\cal F})$ by
$\|T\|=\wedge\{\xi\in L^{0}_{+}({\cal
F})~|~\|Tx\|\leq\xi\|x\|,\forall x\in E\},\forall T\in B(E,F)$, then
$(B(E,F),\|\cdot\|)$ is an $RN$ sapce over $K$ with base
$(\Omega,{\cal F},P)$.}

\th{Proposition 2.4.4 {\rm \cite{Guotx-phd,Guotx-modulehom}}.} {\it
Let $E$ and $F$ be any two $RN$ modules over $K$ with base
$(\Omega,{\cal F},P)$. Then a linear operator $T:E\rightarrow F$ is
a.s. bounded iff $T$ is a continuous module homomorphism from
$(E,{\cal T}_{\epsilon,\lambda})$ to $(F,{\cal
T}_{\epsilon,\lambda})$, and at which case
$\|T\|=\vee\{\|Tx\|~|~x\in E$ and $\|x\|\leq 1\}$.}

\th{Remark 2.4.1.}When ${\cal T}_{\epsilon,\lambda}$ is replaced
with ${\cal T}_c$ Proposition 2.4.4 is also true, even the proof is
easier. The proof of Proposition 2.4.4 uses the fact that an
$L^{0}$-convex set $A$ in an $RN$ module over $K$ with base
$(\Omega,{\cal F},P)$ is a.s. bounded (namely $\vee\{\|x\|~|~a\in
A\}\in L^{0}_{+}({\cal F})$) iff $A$ is ${\cal
T}_{\epsilon,\lambda}$-bounded. An interesting observation should be
made: Let $(S,\|\cdot\|)$ be an $RN$ module over $K$ with base
$(\Omega,{\cal F},P)$ and $A$ a subset of $S$, then $A$ is a.s.
bounded iff $A$ is ${\cal T}_c$-bounded (namely for any ${\cal
T}_c$-neighborhood $U$ of the null vector there exists some $\xi\in
L^{0}_{++}({\cal F})$ such that $A\subset\xi U$). This observation
also holds for any random locally convex module $(S,{\cal P})$ with
base $(\Omega,{\cal F},P)$: $A\subset S$ is ${\cal T}_c$-bounded iff
$A$ is a.s. bounded, namely $\vee\{\|x\|~|~a\in A\}\in
L^{0}_{+}({\cal F})$ for each $\|\cdot\|\in {\cal P}$.

\th{Definition 2.4.5 {\rm\cite{Guotx-phd,Guotx-modulehom}}.} {\it
Let $E$ and $F$ be any two $RN$ spaces over $K$ with base
$(\Omega,{\cal F},P)$ and $T\in B(E,F)$. Define
$T^{\ast}:F^{\ast}\rightarrow E^{\ast}$ by
$T^{\ast}f(x)=f(Tx),\forall x\in E$ and $f\in F^{\ast}$, it is very
easy to prove that $T^{\ast}\in B(F^{\ast},E^{\ast})$ and
$\|T^{\ast}\|=\|T\|$. $T^{\ast}$ is called the conjugate operator of
$T$.}

\th{Definition 2.4.6 {\rm \cite{Guotx-phd,Guotx-modulehom}}.} {\it
Let $(S,\|\cdot\|)$ be an $RN$ space. Define $J:S\rightarrow
S^{\ast\ast}:=(S^{\ast})^{\ast}$ by $(J(x))(f)=f(x),\forall f\in
S^{\ast}$ and $x\in E$, then $J$ is random norm preserving. If $J$
is surjective, then $S$ is called random reflexive.}\vspace{1mm}

Making use of Proposition 2.4.4 Guo first proved in
\cite{Guotx-modulehom} that $B(E,F)$ is ${\cal
T}_{\epsilon,\lambda}$-complete if $E$ and $F$ are $RN$ modules such
that $F$ is ${\cal T}_{\epsilon,\lambda}$-complete, and hence
$S^{\ast\ast}$ is always ${\cal T}_{\epsilon,\lambda}$-complete by
noticing $S^{\ast}$ is an $RN$ module for an $RN$ space $S$; then
Guo further proved in \cite{Guotx-modulehom} that $B(E,F)$ always is
${\cal T}_{\epsilon,\lambda}$-complete for any $RN$ spaces $E$ and
$F$ such that $F$ is ${\cal T}_{\epsilon,\lambda}$-complete. The key
step in completing the proof is that Guo observed in
\cite{Guotx-modulehom} the following: if $S$ is only an $RN$ space,
then the embedding mapping $J:S\rightarrow S^{\ast\ast}$ can be used
to generate a ${\cal T}_{\epsilon,\lambda}$-complete $RN$ module
from $S$. Let $M(S)$ be the ${\cal T}_{\epsilon,\lambda}$-closed
submodule generated by $J(S)$ in $S^{\ast\ast}$, then $M(S)$ is a
${\cal T}_{\epsilon,\lambda}$-complete $RN$ module since
$S^{\ast\ast}$ is ${\cal T}_{\epsilon,\lambda}$-complete.

\th{Proposition 2.4.5 {\rm\cite{Guotx-modulehom}}.} {\it Let $E$ and
$F$ be two $RN$ spaces over $K$ with base $(\Omega,{\cal F},P)$ such
that $F$ is ${\cal T}_{\epsilon,\lambda}$-complete. Define
$L:B(E,F)\rightarrow B(M(E),M(F))$ by
$L(T)=T^{\ast\ast}|_{M(E)},\forall\, T\in B(E,F)$, where
$T^{\ast\ast}=(T^{\ast})^{\ast}:E^{\ast\ast}\rightarrow
F^{\ast\ast}$ is the conjugate operator of $T^{\ast}$. Then $L$ is
random norm preserving and $L(B(E,F))$ is a ${\cal
T}_{\epsilon,\lambda}$-closed subspace of $B(M(E),M(F))$. Specially
$B(E,F)$ is ${\cal T}_{\epsilon,\lambda}$-complete.}

\th{Corollary 2.4.1 {\rm \cite{Guotx-modulehom}}.} {\it $S^{\ast}$
is ${\cal T}_{\epsilon,\lambda}$-complete for any $RN$ space $S$.}

When we generalize the idea of random conjugate spaces from $RN$
spaces to random locally convex spaces, historically there are two
notions of a random conjugate space for a random locally convex
space. It turns out that they just correspond to the locally
$L^{0}$-convex topology and the $(\epsilon,\lambda)$-topology,
respectively, in the context of a random locally convex module!

{\th{Definition 2.4.7 {\rm \cite{Guotx-modulehom}}.} {\it Let
$(S,{\cal P})$ be a random locally convex space over $K$ with base
$(\Omega,{\cal F},P)$. A random linear functional $f:S\rightarrow
L^{0}({\cal F},K)$ is called an a.s. bounded random linear
functional of type \uppercase\expandafter{\romannumeral 1} if there
are some $\xi\in L^{0}_+({\cal F})$ and $Q\in{\cal F}({\cal P})$
such that $|f(x)|\leq\xi\|x\|_Q,\forall x\in S$. Denote by
$S^{\ast}_c$ the linear space of a.s. bounded random linear
functionals of type \uppercase\expandafter{\romannumeral 1} on $S$,
similar to Definition $2.4.3$ $S^{\ast}_c$ becomes a left module
over $L^{0}({\cal F},K)$, called the random conjugate space of type
\uppercase\expandafter{\romannumeral 1} of $S$.}

\th{Definition 2.4.8 {\rm \cite{Survey2}}.} {\it Let $(S,{\cal P})$
be a random locally convex space over $K$ with base $(\Omega,{\cal
F},P)$. A random linear functional $f:S\rightarrow L^{0}({\cal
F},K)$ is called an a.s. bounded random linear functional of type
\uppercase\expandafter{\romannumeral 2} on $S$ if there exist some
$\xi\in L^{0}_+({\cal F})$ and $\|\cdot\|\in{\cal P}_{cc}$ such that
$|f(x)|\leq\xi\|x\|,\forall x\in S$ $($see Section $2.2$ for ${\cal
P}_{cc})$. Denote by $S^{\ast}_{\epsilon,\lambda}$ the $L^{0}({\cal
F},K)$-module of a.s. bounded random linear functionals of type
\uppercase\expandafter{\romannumeral 2} on $S$, called the random
conjugate space of type \uppercase\expandafter{\romannumeral 2} of
$S$.}\vspace{1mm}

Proposition 2.4.6 below is essentially the simpler Lemma 2.12 of
\cite{Guotx-Relations}, which not only considerably simplifies the
proof of the Hahn-Banach theorem for $L^{0}$-linear functions but
also makes it easier for people to understand the topological module
characterizations for $S^{\ast}_c$ and
$S^{\ast}_{\epsilon,\lambda}$.

\th{Proposition 2.4.6 {\rm \cite{Guotx-Relations}}.} {\it Let $S$ be
a left module over the algebra $L^{0}({\cal F},K)$ and
$f:S\rightarrow L^{0}({\cal F},K)$ a random linear functional. Then,
we have the following statements$:$

$(1)$ If $K=R$, then $f$ is an $L^{0}$-linear function iff there
exists an $L^{0}$-sublinear function $p:S\rightarrow L^{0}({\cal
F},R)$ such that $f(x)\leq p(x),\forall x\in S;$

$(2)$ If $K=C$, then $f$ is an $L^{0}$-linear function iff there
exists an $L^{0}$-seminorm $p:S\rightarrow L^{0}_+({\cal F})$ such
that $|f(x)|\leq p(x),\forall x\in S$.}\vspace{1mm}

Corollary 2.4.2 below is easily derived from Propositions 2.4.1 and
2.4.6. Corollary 2.4.2 is known as a special case of the main
results of \cite{Breckner-Scheiber,Vuza} and its proof was also
given in \cite{FKV}. But since not every element of $L^{0}({\cal
F},R)$ has a multiplication inverse element, this brings an obstacle
to one step extension in the process of the proof of Corollary
2.4.2, the complicated methods were used in
\cite{FKV,Breckner-Scheiber,Vuza} in order to overcome this
obstacle.

\th{Corollary 2.4.2 {\rm \cite{Breckner-Scheiber,Vuza}}.} {\it Let
$S$ be a left module over the algebra $L^{0}({\cal F},R)$, $M\subset
S$ a submodule, $f:M\rightarrow L^{0}({\cal F},R)$ an $L^{0}$-linear
function and $p:S\rightarrow L^{0}({\cal F},R)$ an $L^{0}$-sublinear
function such that $f(x)\leq p(x),\forall x\in M$. Then there exists
an $L^{0}$-linear function $g:S\rightarrow L^{0}({\cal F},R)$ such
that $g$ extends $f$ and $g(x)\leq p(x),\forall x\in
S$.}\vspace{1mm}

Corollary 2.4.3 below can be easily derived not only from Corollary
2.4.2 but also from Propositions 2.4.3 and 2.4.6.

\th{Corollary 2.4.3 {\rm \cite{Guotx-Relations}}.} {\it Let $S$ be a
left module over the algebra $L^{0}({\cal F},C)$, $M\subset S$ a
submodule, $f:M\rightarrow L^{0}({\cal F},C)$ an $L^{0}$-linear
function and $p:S\rightarrow L^{0}_+({\cal F})$ an $L^{0}$-seminorm
such that $|f(x)|\leq p(x),\forall x\in M$. Then there is an
$L^{0}$-linear function $g:S\rightarrow L^{0}({\cal F},C)$ such that
$g$ extends $f$ and $|g(x)|\leq p(x),\forall x\in S$.}\vspace{1mm}

The following two propositions give the topological
characterizations of $S^{\ast}_c$ and
$S^{\ast}_{\epsilon,\lambda}$, and hence also an equivalent
definition of either of $S^{\ast}_c$ and
$S^{\ast}_{\epsilon,\lambda}$ as given in \cite{Guotx-Relations}.

\th{Proposition 2.4.7 {\rm \cite{Guotx-Relations}}.} {\it Let
$(S,{\cal P})$ be a random locally convex module over $K$ with base
$(\Omega,{\cal F},P)$ and $f:S\rightarrow L^{0}({\cal F},K)$ a
random linear functional. Then $f\in S^{\ast}_c$ iff $f$ is a
continuous module homomorphism from $(S,{\cal T}_c)$ to
$(L^{0}({\cal F},K),{\cal T}_c)$.}

\th{Proposition 2.4.8 {\rm\cite{Survey2,Guo-Zhu}}.} {\it Let
$(S,{\cal P})$ be a random locally convex module over $K$ with base
$(\Omega,{\cal F},P)$ and $f:S\rightarrow L^{0}({\cal F},K)$ a
random linear functional. Then $f\in S^{\ast}_{\epsilon,\lambda}$
iff $f$ is a continuous module homomorphism from $(S,{\cal
T}_{\epsilon,\lambda})$ to $(L^{0}({\cal F},K),{\cal
T}_{\epsilon,\lambda})$.}\vspace{1mm}

From Propositions 2.4.7 and 2.4.8, we can now give the following
topological versions of Corollaries 2.4.2 and 2.4.3.

\th{Proposition 2.4.9{\rm\cite{Guotx-Relations}}.} {\it Let
$(S,{\cal P})$ be a random locally convex module over $K$ with base
$(\Omega,{\cal F},P)$ and $M\subset S$ a submodule. Then we have the
following statements$:$

$(1)$ every continuous module homomorphism from $(M,{\cal T}_c)$ to
$(L^{0}({\cal F},K),{\cal T}_c)$ can be extended to a continuous
module homomorphism from $(S,{\cal T}_c)$ to $(L^{0}({\cal
F},K),{\cal T}_c);$

$(2)$ every continuous module homomorphism from $(M,{\cal
T}_{\epsilon,\lambda})$ to $(L^{0}({\cal F},K),{\cal
T}_{\epsilon,\lambda})$ can be extended to a continuous module
homomorphism from $(S,{\cal T}_{\epsilon,\lambda})$ to $(L^{0}({\cal
F},K),{\cal T}_{\epsilon,\lambda})$.}

\th{Proposition 2.4.10 {\rm \cite{Guotx-Relations}}.} {\it Let
$(S,{\cal P})$ be a random locally convex space. Then
$S^{\ast}_{\epsilon,\lambda}=S^{\ast}_c$ if ${\cal P}$ has the
countable concatenation property $($generally, it is obvious that
$S^{\ast}_c\subset S^{\ast}_{\epsilon,\lambda})$. In particular,
$S^{\ast}_{\epsilon,\lambda}=S^{\ast}_c$ for any $RN$ space
$(S,\|\cdot\|)$.}

\th{Remark 2.4.2.} Before 1995, the focus of our work is on $RN$
spaces and indeed we also obtained several pleasant results, for
example, Proposition 2.4.3, Proposition 2.4.5 and Corollary 2.4.1.
But the results in the paper \cite{Guotx-extension} and
 further in \cite{Guotx-modulehom} (for example, Proposition 2.4.4) made us realize the fundamental importance of the module structure
  of an $RN$ module, thus after 1995 the theory of $RN$ modules together with their random conjugate spaces has been our concern.

\section{Representation theorems of random conjugate spaces}

\th{Proposition 3.1 {\rm(Riesz's representation theorem)
\cite{Guo-You}}.} {\it Let $(S,\langle\cdot,\cdot\rangle )$ be a
$\cal T_{\epsilon,\lambda}$-complete $RIP$ module. Then there exists
a unique $\pi(f)$ in $S$ for each $f\in
 S^{\ast}_{\epsilon,\lambda}$ such that $f(x)=\langle x,\pi(f)\rangle , \forall x\in
 S$, and $\|{\pi(f)}\|=\|f\|$.}

\th{Corollary 3.1 {\rm\cite{Guotx-Relations}}.} {\it Let $(S,\langle
\cdot,\cdot\rangle )$ be a ${\cal T}_{c}$-complete $RIP$ module such
that $S$ has the countable concatenation property. Then there exists
a unique $\pi(f)$ in $S$ for each $f\in S^\ast_c$ such that
$f(x)=\langle x,\pi(f)\rangle , \forall x\in
 S$, and $\|{\pi(f)}\|=\|f\|$.}\vspace{1mm}

Proposition 3.1 is essential and Corollary 3.1 is merely a
 consequence of Propositions 3.1, 2.4.10 and
 2.2.6. Here, we would like to review the proof of Proposition\,\,3.1.
 First, its proof is considerably different from
that of the classical Riesz's representation theorem in Hilbert
Spaces and the classical case only needs to utilize the orthogonal
decomposition theorem, whereas the proof of Proposition 3.1 forces
us to work out a countable concatenation technique in order to
obtain $\pi(f)$ by means of the countable concatenation property of
$S$ under $\cal T_{\epsilon,\lambda}$, and thus one should not
surprise at the hypothesis on Corollary 3.1. Secondly, the following
orthogonal decomposition theorem is, of course, used in the proof of
Proposition 3.1.

\th{Proposition 3.2 {\rm \cite{Guo-You,Guotx-somebasic}}.} {\it Let
$(S,\langle \cdot,\cdot\rangle )$ be a $\cal
T_{\epsilon,\lambda}$-complete $RIP$ module over $K$ with base
$(\Omega,\cal F,P)$, $M\subset S$ a $\cal
T_{\epsilon,\lambda}$-closed submodule and $M^\bot=\{x\in
S~|~\langle x,y\rangle =0, \forall y\in M\}$. Then $S=M\oplus
M^\bot$.}

\pf{Proof.} Let $x$ be any element in $S$, we prove that there
exists a unique $x_0\in M$ such that $x-x_0\in M^\bot$. First, let
$d(x,M)=\wedge \{\|x-y\|~|~y\in M\}$, and for any $y_1$, $y_2\in M$
let $A=[\|x-y_1\|\leq \|x-y_2\|]$, where $A$ is the equivalence
class of $A^0$, $A^0=\{\omega \in \Omega~|~\|x-y_1\|^0(\omega)\leq
\|x-y_2\|^0(\omega)\}$, $\|x-y_1\|^0$ and $\|x-y_2\|^0$ are
arbitrarily chosen representatives, respectively, and
$I_A:=\tilde{I}_{A^0}$. Then one can easily check that
$y_3=I_Ay_1+(1-I_A)y_2\in M$ and satisfies the relation:
$\|x-y_3\|=\|x-y_1\|\wedge \|x-y_2\|$, which shows that
$\{\|x-z\|~|~z\in M\}$ is directed downwards. Then by Proposition
1.3.5 there exists a sequence $\{z_n, n\in N\}$ in $M$ such that
$\{\|x-z_n\|~|~n\in N\}$ converges to $d(x,M)$ in a nonincreasing
manner. One can prove that $\{z_n, n\in N\}$ is a $\cal
T_{\epsilon,\lambda}$-Cauchy sequence completely similar to the
classical case, hence convergent to some point $x_0$. Further, one
can also verify that  $x_0$ is just as desired as in the classical
case.

As we have seen, the proof of Proposition 3.2 is only a copy of the
proof of its classical prototype, but as a spacial case of
Proposition 3.2, Corollary 3.2 below used to appear  in the
mathematical finance literature
\cite{Follmer-Schied-Stocha,Schachermayer}, but where its proof is
indirect by a technique of converting the orthogonal decomposition
problem in a special ${\cal T}_{\epsilon,\lambda}$-complete $RIP$
module to the corresponding problem in a Hilbert space. Now, we can
give a straightforward proof.

\th{Corollary 3.2 {\rm\cite{Follmer-Schied-Stocha,Schachermayer}}.}
{\it Let $(\Omega,{\cal F}_1,P)$ be a probability space, ${\cal
F}_0$ a sub $\sigma$-algebra of ${\cal F}_1$, $y$ in $L^0({\cal
F}_1,R^d)$ $($where $R^d$ is the $d$-dimensional Euclidean space$)$,
$N=\{x\in L^0({\cal F}_0,R^d)~|~\langle x,y\rangle=0\}$ and
$N^\bot=\{z\in L^0({\cal F}_0,R^d)~|~\langle x,z\rangle =0, \forall
x\in N\}$. Then $L^0({\cal F}_0,R^d)=N\oplus N^\bot$.}

\pf{Proof.}Take $S=L^0({\cal F}_0,R^d)$ and $M=N$ in Corollary 3.2,
then the desired follows.

\th{Definition 3.1.} {\it Let $E$ and $F$ be any two $RN$ modules
over $K$ with base $(\Omega,{\cal F},P)$. A mapping $T:E\rightarrow
F$ is called an isometric isomorphism between $E$ and $F$ if $T$ is
a random norm preserving module isomorphism.}\vspace{1mm}

Let $L^0({\cal F},B)$ and $L^0({\cal
F},B^{\prime},\textmd{w}^{\ast})$ be the same as in Section 2.3. For
any $x$ in $L^0({\cal F},B)$ and $y$ in $L^0({\cal
F},B^{\prime},\textmd{w}^{\ast})$, let $x^0$ and $y^0$ be any chosen
representatives of $x$ and $y$, respectively, $\langle
x^0,y^0\rangle : \Omega\rightarrow K$ is defined by $\langle
x^0,y^0\rangle (\omega)=(y^0(\omega))(x^0(\omega)),\forall \omega
\in \Omega$. Then it is clear that $\langle x^0,y^0\rangle $ is a
$K$-valued $\cal F$-random variable on $(\Omega,{\cal F},P)$, and
denote by $\langle x,y\rangle $ the equivalence class of $\langle
x^0,y^0\rangle $. Now, for each $y\in L^0({\cal
F},B^{\prime},\textmd{w}^{\ast})$, define $T(y): L^0({\cal
F},B)\rightarrow L^0({\cal F},K)$ by $(T(y))(x)=\langle x,y\rangle ,
\forall x\in L^0({\cal F},B)$. Then we have the following:

\th{Proposition 3.3 {\rm\cite{Guotx-radon}}.} {\it Let
$(\Omega,{\cal F},P)$ be a complete probability space. Then $T$
defined as above is an isometric isomorphism from $L^0({\cal
F},B^{\prime},\textmd{w}^{\ast})$ onto $(L^0({\cal F},B))^\ast$
$($namely the random conjugate space of $L^0({\cal
F},B))$.}\vspace{1mm}

Since $L^0({\cal F},B^\prime)$ can be regarded as a submodule of
$L^0({\cal F},B^{\prime},\textmd{w}^{\ast})$, we should consider the
following problem: When is the restriction of $T$ as in Proposition
3.3 to $L^0({\cal F},B^\prime)$ also an isometric isomorphism
between $L^0({\cal F},B^\prime)$ and $(L^0({\cal F},B))^\ast$ ? We
have the following answer:

\th{Proposition 3.4 {\rm\cite{Guotx-radon}}.} {\it $T$ is an
isometric isomorphism between $L^0({\cal F},B^\prime)$ and
$(L^0({\cal F},B))^\ast$ iff $B^\prime$ has the Radon-Nikod\'{y}m
property with respect to $(\Omega,{\cal F},P)$.}

\th{Corollary 3.3 {\rm\cite{Guotx-radon}}.} {\it Let $(\Omega,{\cal
F},P)$ be a complete probability space. Then $B^\prime$ has the
Radon-Nikod\'{y}m property with respect to $(\Omega,{\cal F},P)$ iff
there is a $B^\prime$-valued random variable $\bar{q}$ for each
$B^\prime$-valued $\textmd{w}^{\ast}$-random variable $q$ such that
$q$ and $\bar{q}$ are $\textmd{w}^\ast$-equivalent to each
other.}\vspace{1mm}

Proposition 3.5 below provides a pleasant connection between the
random conjugate space $S^\ast$ of an $RN$ module $S$ and the
classical conjugate space $(L^p(S))^\prime$ of $(L^p(S))$, which
enables us to establish many difficult results, for example,
Proposition 3.6 below, all the results in Section 4, and the main
result of \cite{Guo-Xiao-Chen}.

\th{Proposition 3.5 {\rm \cite{Guotx-ref,Survey2}}.} {\it Let
$(S,\|\cdot\|)$ be an $RN$ module over $K$ with base $(\Omega,{\cal
F},P)$ and $1\leq p<\infty$. Then
$T:(L^q(S^\ast),\|\cdot\|_q)\rightarrow (L^p(S))^\prime$ is an
isometric isomorphism. Where $q$ is the H\"{o}lder conjugate number
of $p$, $L^p(S)$ and $L^q(S^\ast)$ are understood as in Section
$2.2$, and for each $f\in L^q(S^\ast)$, $T(f): L^p(S)\rightarrow K$
is defined by $(T(f))(g)=\int_\Omega f(g)dP, \forall g\in L^p(S)$.}

\th{Proposition 3.6 {\rm\cite{Guotx-Relations}}.} {\it Let
$(S,\|\cdot\|)$ be an $RN$ module over $K$ with base $(\Omega,{\cal
E},P)$, $\cal F\subset\cal E$ a sub $\sigma$-algebra and $1\leq
p<\infty$ with H\"{o}lder conjugate number $q$. Then $T: (L^q_{\cal
F}(S^\ast),|||\cdot|||_q)\rightarrow (L^p_{\cal F}(S))^\ast$ is
isometric isomorphism, where $L^p_{\cal F}(S)$ and $L^q_{\cal
F}(S^\ast)$ are understood as in Section $2.3$, and for each $f\in
L^{q}_{\cal F}(S^{\ast})$ $T(f):L^{p}_{\cal F}(S)\rightarrow
L^{0}({\cal F}, K)$ is defined by $(T(f))(g)=E(f(g)~|~{\cal F})$,
$\forall g \in L^{p}_{\cal F}(S)$.}

\th{Corollary 3.4 {\rm\cite{KV}}.} {\it Let $(\Omega,{\cal E},P)$ be
a probability space, $\cal F \subset \cal E$ a sub $\sigma$-algebra
and $1\leqslant p<+\infty$ with the H\"older conjugate number $q$.
Then $T:L^{q}_{\cal F}({\cal E})\rightarrow (L^{p}_{\cal F}({\cal
E}))^{\ast}$ is an isometric isomorphism, where for each $f \in
L^{q}_{\cal F}({\cal E})$, $T(f):L^{p}_{\cal F}({\cal E})\rightarrow
L^{0}({\cal F}, R)$ is defined by $(T(f))(g)=E(f\cdot g ~|~ {\cal
F})$, $\forall g \in L^{p}_{\cal F}({\cal E})$.}

\th{Remark 3.1.} The original proof of Corollary 3.4 in \cite{KV}
only shows that $T$ is a module isomorphism, whereas the isometric
property of $T$ was proved in \cite{Guotx-Relations} as a special
case of Proposition 3.6. Corollary 3.4 is crucial in the dual
representation of conditional risk measures; see Section 7 for
details.

\section{Characterization for random reflexivity}

According to Definition 2.4.6, if an $RN$ space is random reflexive
then it has to be a both ${\cal T}_{c}$-complete and ${\cal
T}_{\epsilon,\lambda}$-complete $RN$ module with the countable
concatenation property since $S^{\ast\ast}$ has all the properties.
In fact, as analyzed in \cite{Guotx-Relations}, random reflexivity
is independent of the special choice of two kinds of topologies, and
hence we will not mention the topologies in this section.

\th{Proposition 4.1 {\rm\cite{Guotx-radon}}.} {\it $L^{0}({\cal F},
B)$ is random reflexive iff $B$ is reflexive.}

\th{Proposition 4.2 {\rm\cite{Guotx-ref}}.} {\it An $RN$ module $S$
is random reflexive iff $L^{p}(S)$ is reflexive for any given $p$
such that $1<p<+\infty$.}

\th{Proposition 4.3 {\rm(The James Theorem) \cite{Guo-Li}}.} {\it A
complete $RN$ module $S$ is random reflexive iff there exists $x \in
S(1)$ for each $f \in S^{\ast}$ such that $f(x)=\|f\|$, where
$S(1)=\{y \in S~|~ \|y\|\leqslant 1\}$.}

\th{Proposition 4.4 {\rm\cite{Guotx-compre}}.} {\it Let $(S,
\|\cdot\|)$ be an $RN$ module over $K$ with base $(\Omega,{\cal
E},P)$, $\cal F \subset \cal E$ a sub $\sigma$-algebra and
$1<p<+\infty$. Then $L^{p}_{\cal F}(S)$ is random reflexive iff $S$
is random reflexive. Specially, $L^{p}_{\cal F}({\cal E})$ is random
reflexive.}

\section{Hyperplane separation theorems}

For a random locally convex module $(E,{\cal P})$ over $K$ with base
$(\Omega,{\cal F},P)$, $x \in E$ and $G \subset E$ a subset. For
each $Q \in {\cal F}({\cal P})$, let $d^{\ast}_{Q} (x,G)=
\wedge\{\|x-y\|_{Q}~|~y\in G\}$, and
$d^{\ast}(x,G)=\vee\{d^{\ast}_{Q} (x,G)~|~Q\in {\cal F}({\cal
P})\}$. Then $d^{\ast}(x,G) \in \bar{L}^{0}_{+}({\cal F})$ in
general.

\th{Proposition 5.1 {\rm\cite{Guotx-Relations,Guo-Xiao-Chen}}.} {\it
Let $(E,{\cal P})$ be a random locally convex module over $K$ with
base $(\Omega,{\cal F},P)$, $x \in E$, $G$ a nonempty ${\cal
T}_{\epsilon,\lambda}$-closed $L^{0}$-convex subset of $E$ such that
$x \notin G$, and $\xi$ a chosen representative of $d^{\ast}(x,G)$.
Then there exists an $f \in E^{\ast}_{\epsilon,\lambda}$ such that
the following are satisfied$:$

$(1)$ $(Ref)(x)>\vee\{(Ref)(y)~|~y\in G\};$

$(2)$ $(Ref)(x)>\vee\{(Ref)(y)~|~y\in G\}$ on $\{\xi>0\}$.}

\th{Proposition 5.2 {\rm\cite{Guotx-Relations}}.} {\it Let $(E,{\cal
P})$ be a random locally convex module over $K$ with base
$(\Omega,{\cal F},P)$ such that $\cal P$ has the countable
concatenation property, $x \in E$ and $G$ a nonempty ${\cal
T}_{c}$-closed $L^{0}$-convex subset of $E$ such that $x \notin G$
and $G$ has the countable concatenation property. Then there exists
an $f \in E^{\ast}_c$ such that the following are satisfied$:$

$(1)$ $(Ref)(x)>\vee\{(Ref)(y)~|~y\in G\};$

$(2)$ $(Ref)(x)>\vee\{(Ref)(y)~|~y\in G\}$ on $\{\xi>0\}$,

\noindent where $\xi$ is any chosen representative of
$d^{\ast}(x,G)$.}

\th{Corollary 5.1 {\rm\cite{Guotx-Relations,FKV}}.} {\it Let
$(E,{\cal P})$ be a random locally convex module over $K$ with base
$(\Omega,{\cal F},P)$ such that $\cal P$ has the countable
concatenation property, $x \in E$ and $G\subset E$ a nonempty ${\cal
T}_{c}$-closed $L^{0}$-convex subset with the countable
concatenation property. If $\tilde{I}_{A}\{x\}\cap
\tilde{I}_{A}G=\emptyset$ for all $A \in {\cal F}$ with $P(A)>0$,
then there exists an $f \in E^{\ast}_c$ and $\varepsilon \in
L^{0}_{++}({\cal F})$ such that $(Ref)(x)>(Ref)(y)+\varepsilon$ on
$\Omega$ for all $y \in G$.}

\th{Remark 5.1.} Corollary 5.1 improves Theorem 2.8 of \cite{FKV} in
that $G$ is assumed to have the countable concatenation property,
whereas Theorem 2.8 of \cite{FKV} did not make the hypothesis. In
fact, we recently constructed a counterexample in
\cite{Guo-Zhao-Zeng} which shows that both Theorem 2.8 and Lemma
2.28 of \cite{FKV} may be not true if the hypothesis is removed.

Now, we can give the improved version of \cite[Lemma
2.28]{FKV}\,---\,Proposition 5.3 below, which is very useful in
improving the main results in \cite{FKV}, see Section 7.

\th{Proposition 5.3 {\rm\cite{Guo-Zhao-Zeng}}.} {\it Let $(E,{\cal
P})$ be a random locally convex module over $K$ with base
$(\Omega,{\cal F},P)$ such that $\cal P$ has the countable
concatenation property, $x \in E$ and $G\subset E$ a ${\cal
T}_{c}$-closed subset with the countable concatenation property. If
$\tilde{I}_{A}\{x\}\cap \tilde{I}_{A}G=\emptyset$ for all $A \in
{\cal F}$ with $P(A)>0$, then there exists an $L^{0}$-convex,
$L^{0}$-absorbent and $L^{0}$-balanced ${\cal T}_{c}$- neighborhood
$U$ of $0 \in E$ such that $\tilde{I}_{A}(x+U)\cap
\tilde{I}_{A}(G+U)=\emptyset$ for all $A \in {\cal F}$ with
$P(A)>0$.}

\th{Proposition 5.4 {\rm \cite{FKV}}.} {\it Let $(E,{\cal P})$ be a
random locally convex module over $K$ with base $(\Omega,{\cal
F},P)$, $G$ and $M$ two $L^{0}$-convex subsets of $E$ such that $G$
is also nonempty and ${\cal T}_{c}$-open. If $\tilde{I}_{A}G\cap
\tilde{I}_{A}M=\emptyset$ for all $A \in {\cal F}$ with $P(A)>0$,
then there exists an $f \in E^{\ast}_c$ such that $(Ref)(x)<
(Ref)(y)$ on $\Omega$ for all $x \in G$ and $y \in M$.}

\th{Remark 5.2.} Proposition 5.4 is peculiar to the locally
$L^0$-convex topology ${\cal T}_{c}$ since ${\cal
T}_{\epsilon,\lambda}$ is too weak to ensure the existence of a proper,
nonempty $L^{0}$-convex and ${\cal T}_{\epsilon,\lambda}$-open
subset in a random locally convex module. By the way, we proved in
\cite{Guotx-Relations} that Proposition 5.1 implies both Proposition
5.2 and Corollary\,\,5.1.

\section{Random duality with respect to the locally
$L^{0}$-convex topology}

The theory of a random locally convex module $(E,{\cal P})$ is
considerably different from the theory of an ordinary locally convex
space in that $\mathcal{P}$ can induce the two kinds of
topologies\,---\,the locally $L^{0}$-convex topology ${\cal T}_{c}$
and the $(\epsilon,\lambda)$-topology $\cal
T_{\epsilon,\lambda}$.Thus the theory of random duality based on the
framework of a random locally convex module should have two
kinds\,---\,corresponding to the above two kinds of topologies,
respectively.

The theory of random duality corresponding to the
$(\epsilon,\lambda)$-topology was presented and studied in
\cite{Guotx-randomduality,Survey2,Guo-Chen} where we could only
speak of random compatible structure and random admissible structure
(they were defined as a family of $L^{0}$-seminorms, respectively)
rather than random compatible topology and random admissible
topology because what really plays a crucial role in random duality
is a family of $L^{0}$-seminorms. Thanks to the contribution made by
Filipovi\'c et al. in \cite{FKV}, namely any locally $L^{0}$-convex
topology can also be induced by a family of $L^{0}$-seminorms, which
enables us to speak of random compatible and random admissible
topologies under the framework of locally $L^{0}$-convex modules
(namely under the framework of locally $L^{0}$-convex topologies).

The purpose of this section is to develop the theory of random
duality under the locally $L^{0}$-convex topology. All the results
in this section without mention of reference belong to the author,
which together with some other interesting results will be published
in our forthcoming joint paper \cite{Guo-Zhao-Zeng}.

Since we will consider more than one family of $L^{0}$-seminorms on
a given $L^{0}(\mathcal {F},K)$-module $E$. Given a family $\mathcal
{P}$ of $L^{0}$-seminorms on the $L^{0}(\mathcal {F},K)$-module such
that $(E,{\cal P})$ become a random locally convex module, in this
section we always use $\mathcal {P}_c$ and $\mathcal
{P}_{\epsilon,\lambda}$ rather than $\mathcal{T}_c$ and
$\mathcal{T}_{\epsilon,\lambda}$ as in the other sections for the
locally $L^{0}$-convex topology and the
$(\epsilon,\lambda)$-topology induced by $\mathcal {P}$,
respectively. At the same time we always use $( {E},\mathcal
{P})_c^*$ for the random conjugate space consisting of all
continuous module homomorphisms from $({E},\mathcal {P}_c)$ to
$(L^{0}(\mathcal {F},K),|\cdot|_c)$ (namely previous $E_c^*$), and
$({E},\mathcal {P})_{\epsilon,\lambda}^*$ for the random conjugate
space consisting of all continuous module homomorphism from
$({{E},\mathcal {P}_{\epsilon,\lambda}})$ to $(L^{0}(\mathcal {F},
{K}),|\cdot|_{\epsilon,\lambda})$ (namely previous
$E_{\epsilon,\lambda}^*$), where $|\cdot|$ denotes the $L^{0}$-norm
on $L^{0}(\mathcal {F},K)$, and $|\cdot|_c$ and
$|\cdot|_{\epsilon,\lambda}$ denote the locally $L^{0}$-convex
topology and $(\epsilon,\lambda)$-topology on $L^{0}(\mathcal {F},
{K})$, respectively.

By the way, to contrast with the results obtained in \cite{Guo-Chen},
we will mention some results of \cite{Guo-Chen} in some places of
this section in time.

\subsection{Random compatible topology}

\th{Definition 6.1.1 {\rm\cite{Guotx-randomduality}}.} {\it Let $E$
and $F$ be two left modules over the algebra $L^{0}(\mathcal
{F},{K})$ and the mapping $\langle\cdot,\cdot\rangle$ : $E\times
F\rightarrow L^{0}(\mathcal {F},{K})$ a bi-module homomorphism. Then
$E$ and $F$ are called a pair in random duality with respect to
$\langle\cdot,\cdot\rangle$ over $K$ with base $(\Omega,\mathcal
{F},P)$ if the following axioms are satisfied$:$

$(1)$ $\langle x,y\rangle=0$ for each $y\in F$ iff $x=0;$

$(2)$ $\langle x,y\rangle=0$ for each $x\in E$ iff
$y=0$.}\vspace{1mm}

For the sake of convenience,we also say that $\langle E,F\rangle$ is
a random duality pair over $K$ with base $(\Omega,\mathcal {F},{P})$
if $E$, $F$ and $\langle\cdot,\cdot\rangle$ satisfy the above two
conditions.

One can easily find that the notion of the left (right) regularity
of a random duality pair $\langle E,F\rangle$ as introduced in
\cite{Guo-Chen} is equivalent to saying that $E$ (resp. $F$) has the
countable concatenation property.

Given a random duality pair $\langle E,F\rangle$ over $K$ with base
$(\Omega,\mathcal {F},{P})$, $\sigma{(E,F)}$ always denotes the
family $\{{\|\cdot\Arrowvert}_f:f\in F\}$ of $L^{0}$-seminorms on
$E$, where ${{\|\cdot\Arrowvert}_f:E\rightarrow L_+^{0}(\mathcal
{F})}$ is defined by ${\|e\Arrowvert}_f=|\langle e,f\rangle
\arrowvert$,$\forall e\in E$.

To establish the representation theorem of the random conjugate
space $(E,\sigma (E,F))_c^*$, let us first recall a piece of linear
functionals (see \cite[Theorem 21.17]{Berberian}): let $E$ be a
linear space over $K$, $f_1,f_2,\ldots,f_n$ and $g$ linear
functionals on $E$, then there are
$\alpha_1,\alpha_2,\ldots,\alpha_n$ in $K$ such that
$g=\sum_{k=1}^{n}\alpha_k f_k$ iff $\bigcap_{k=1}^nN(f_k)\subset
N(g)$, where $N(f)$ stands for the null space of a linear functional
$f$ on $E$. When we generalized the classical result to
$L^{0}$-linear functions on an $L^{0}(\mathcal {F}, {K})$-module $E$
in \cite{Guotx-randomduality}, we again made use of the countable
concatenation property of $L^{0}(\mathcal {F},{K})$, and thus this
generalization is not trivial, as shown in \cite{Guo-Chen}.

\th{Proposition 6.1.1 {\rm\cite{Guotx-randomduality,Guo-Chen}}.}
{\it Let $E$ be a left module over the algebra $L^{0}(\mathcal
{F},{K})$, $f_1,f_2,\ldots,f_n$ and $g$ $L^{0}$-linear functions on
$E$. Then there exist $\xi_1,\xi_2,\ldots,\xi_n$ in $L^{0}({\cal
F},K)$ such that $g=\sum_{k=1}^n\xi_k f_k$ iff
$\bigcap_{k=1}^nN(f_k)\subset N(g)$.}\vspace{1mm}

Proposition 6.1.2 below was first obtained in
\cite{Guotx-randomduality} because the very random conjugate space
$E_c^*$ was employed in \cite{Guotx-randomduality}.

\th{Proposition 6.1.2 {\rm\cite{Guotx-randomduality}}.} {\it Let
$\langle E,F\rangle$ be a random duality pair over $K$ with base
$(\Omega, \cal F,P)$. Then $(E,\sigma (E,F))_c^*=F$, namely there
exists a unique $y$ in $F$ for each $f\in (E,\sigma (E,F))_c^*$ such
that $f(x)=\langle x,y\rangle$, $\forall x\in E$.}

\th{Remark 6.1.1.} In \cite{Guo-Chen}, we proved that for each $f\in
(E,\sigma (E,F))_{\epsilon,\lambda}^*$ there exist a countable
subset $\{y_n~|~n\in N\}$ in $F$ and a countable partition
$\{A_n~|~n\in N\}$ of $\Omega$ to $\cal F$ such that $f(x)=\sum
_{n\geq 1}\tilde{I}_{A_n}\langle x,y_n\rangle$, $\forall x \in E$.

\th{Definition 6.1.2.} {\it Let $\langle E,F\rangle$ be a random
duality pair over $K$ with base $(\Omega,\cal F,P)$. A Hausdorff
locally $L^{0}$-convex topology $\cal {T}$ for $E$ $($namely
$(E,\cal T)$ forms a Hausdorff locally $L^{0}$-convex module$)$ is
called a random compatible topology with $\langle E,F \rangle$ if
$E_c^{*}=F$.}

\th{Remark 6.1.2.}In Definition 6.1.2, $E_c^{*}$ is exactly the
$L^{0}$-module of continuous module homomorphisms from $(E,\cal T)$
to $(L^{0}({\cal F},K),|\cdot|_c)$. In \cite{Guo-Chen}, we say that
a family $\cal P$ of $L^{0}$-seminorms on $E$ is a random compatible
structure with $\langle E,F\rangle$ if $F$ has the countable
concatenation property, $(E,\cal P)$ is a random locally convex
module and $(E,{\cal P})_{\epsilon,\lambda}^*=F$.

\th{Proposition 6.1.3 {\rm(Mackey topology)}.} {\it Let $\langle
E,F\rangle$ be a random duality pair. Then there is the greatest
random compatible topology for $E$ with $\langle E,F\rangle$, called
random Mackey topology.}

\th{Definition 6.1.3 {\rm\cite{Guo-Chen}}.} {\it Let $\langle
E,F\rangle$ be a random duality pair, $A\subset E$ and $B\subset F$.
$A^{0}=\{y\in F~|~|\langle x,y\rangle|\leq 1,\forall x\in A\}$ is
called the polar of $A$, and $B^{0}=\{x\in E~|~|\langle
x,y\rangle|\leq 1,\forall y\in B\}$ is called the polar of $B$.}

\th{Definition 6.1.4.} {\it Let $(E,{\cal T})$ be a locally
$L^{0}$-convex module over $K$ with base $(\Omega,{\cal F},P)$ and
$A\subset E$. Then $A$ is ${\cal T}$-bounded if $A$ can be
$L^{0}$-absorbed by every neighborhood $U$ of $0\in E$ $($namely
there exists some $\xi\in L^{0}_{++}({\cal F})$ such that
$A\subset\xi U)$.}

\th{Proposition 6.1.4.} {\it Let $(E,{\cal T})$ be a locally
$L^{0}$-convex module over $K$ with base $(\Omega,{\cal F},P)$ and
$A\subset E$. Then $A$ is ${\cal T}$-bounded iff $\vee\{\|a\|~|~a\in
A\}\in L^{0}_{+}({\cal F})$ for every $\|\cdot\|\in{\cal P}$, where
${\cal P}$ is a family of $L^{0}$-seminorms on $E$ which generates
${\cal T}$.}\vspace{1mm}

Given a locally $L^{0}$-convex module $(E,{\cal T})$ over $K$ with
base $(\Omega,{\cal F},P)$, $E^{\ast}_c$ denotes the random
conjugate space of $(E,{\cal T})$, namely $E^{\ast}_c$ denotes the
$L^{0}$-module of continuous module homomorphisms from $(E,{\cal
T})$ to $(L^{0}({\cal F},K),|\cdot|_c)$. Then $\langle
E,E^{\ast}_c\rangle$ forms a random duality pair over $K$ with base
$(\Omega,{\cal F},P)$ with $\langle \cdot,\cdot\rangle:E\times
E^{\ast}_c\rightarrow L^{0}({\cal F},K)$ defined by $\langle
g,f\rangle=f(g),\forall (g,f)\in E\times E^{\ast}_c$. As usual, we
briefly use $\sigma_c(E,E^{\ast})$ for $\sigma_c(E,E_c^{\ast})$.

Proposition 6.1.4 shows that the notion of a ${\cal T}$-bounded set
is equivalent to that of an a.s. bounded set in terms of
\cite{Guotx-modulehom}, and thus a result of \cite{Guotx-modulehom}
has implied the following:

\th{Proposition 6.1.5 {\rm\cite{Guotx-modulehom}}.} {\it Let
$(E,{\cal T})$ be a Hausdorff locally $L^{0}$-convex module over $K$
with base $(\Omega,{\cal F},P)$ and $A\subset E$. Then $A$ is ${\cal
T}$-bounded iff $A$ is $\sigma_c(E,E^{\ast})$-bounded, namely $f(A)$
is $|\cdot|_c$-bounded in $(L^{0}({\cal F},K),|\cdot|_c)$ for each
$f\in E_c^{\ast}$.}

\th{Definition 6.1.5.} {\it Let $(E,{\cal T})$ be a Hausdorff
locally $L^{0}$-convex module such that $S$ has the countable
concatenation property. For a subset $A$ of $E$, then the set ${\cal
T}-BC_4(A)=\cap\{G\subset E~|~G\supset A$ and $G$ is an
$L^{0}$-balanced, $L^{0}$-convex and ${\cal T}$-closed set with the
countable concatenation property$\}$ is called the $L^{0}$-balanced,
$L^{0}$-convex, ${\cal T}$-closed countable concatenation hull of
$A$.}

\th{Proposition 6.1.6 {\rm (Random bipolar theorem)}.} {\it Let
$\langle E,F\rangle$ be a random duality pair such that $E$ has the
countable concatenation property. Then ${\cal T}-BC_4(A)=A^{00}$ for
each subset $A$ of $E$ and for each random compatible topology of
${\cal T}$.}

\th{Remark 6.1.3.}The reason why random bipolar theorem is so
complicated is that its proof needs the use of the hyperplane
separation theorem under the locally $L^{0}$-convex
topology\,---\,Proposition 5.2, so that the reader can easily see
why we need to consider the countable concatenation operations
twice. Random bipolar theorem under the
$(\epsilon,\lambda)$-topology is closer to the classical bipolar
theorem, see \cite[Theorem\,\,3.4]{Guo-Chen}.

\subsection{Random admissible topology}

\th{Definition 6.2.1.} {\it Let $\langle E,F\rangle$ be a random
duality pair over $K$ with base $(\Omega,{\cal F},P)$ and ${\cal A}$
a family of $\sigma_c(F,E)$-bounded subsets of $F$. For each $A\in
{\cal A}$, let $\|x\|_A=\vee\{|\langle x,y\rangle|~|~y\in
A\},\forall x\in E$ $($$\|\cdot\|_A$ is well defined by Proposition
$6.1.4)$. Then the locally $L^{0}$-convex topology induced by the
family ${\cal P}:=\{\|\cdot\|_A:A\in{\cal A}\}$ of $L^{0}$-seminorms
on $E$, denoted by ${\cal T}_{\cal A}$, is called the random uniform
convergence topology of $E$ over ${\cal A}$. Furthermore, if
$(E,{\cal T}_{\cal A})^{\ast}_c\supset F$ then ${\cal T}_{\cal A}$
is called a random admissible topology of $E$ with respect to
$\langle E,F\rangle$.}

\th{Proposition 6.2.1.} {\it Let $\langle E,F\rangle$, ${\cal A}$
and ${\cal T}_{\cal A}$ be the same as in Definition $6.2.1.$ Then
${\cal T}_{\cal A}$ is Hausdorff iff $\cup{\cal A}$ is total for
$E$, namely $\langle x,y\rangle=0~\forall y\in \cup{\cal A}$ implies
$x=\theta$, in turn iff $span({\cal A}):=$ the submodule generated
by $\cup{\cal A}$, is $\sigma_{\epsilon,\lambda}(F,E)$-dense in
$F$.}

\th{Definition 6.2.2 {\rm\cite{FKV}}.} {\it Let $(E,{\cal T})$ be a
locally $L^{0}$-convex module over $K$ with base $(\Omega,{\cal
F},P)$. An $L^{0}$-seminorm $\|\cdot\|:E\rightarrow L^{0}_{+}({\cal
F})$ is called ${\cal T}$-lower semicontinuous if for each $\xi\in
L^{0}_{+}({\cal F})$ the set $\{x\in E~|~\|x\|\leq\xi\}$ is ${\cal
T}$-closed.}

\th{Proposition 6.2.2.} {\it Let $\langle E,F\rangle$ be a random
duality pair such that $E$ has the countable concatenation property.
Then a locally $L^{0}$-convex topology ${\cal T}$ for $E$ is a
random admissible topology iff ${\cal T}$ satisfies the following
two conditions$:$

$(1)$ ${\cal T}\supset\sigma_c(E,F);$

$(2)$ ${\cal T}$ is induced by a family of $\sigma_c(E,F)$-lower
semicontinuous $L^{0}$-seminorms on $E$.

Besides, the above $(2)$ is equivalent to the following$:$

$(3)$ There is a neighborhood base ${\cal U}$ of $0\in E$ for ${\cal
T}$ such that each $U\in{\cal U}$ is an $L^{0}$-convex,
$L^{0}$-balanced, $L^{0}$-absorbent and $\sigma_c(E,F)$-closed set
with the countable concatenation property.}

\th{Proposition 6.2.3.}\quad{\it Let $(E,{\cal T})$ be a locally
$L^{0}$-convex module over $K$ with base $(\Omega,{\cal F},P)$ such
that $E$ has the countable concatenation property. Suppose ${\cal
E}$ is the family of ${\cal T}$-equicontinuous subsets of
$E^{\ast}_c$, then ${\cal T}= {\cal T}_{\cal E}$, and hence ${\cal
T}$ is a random admissible topology of $E$ with respect to the
natural pair $\langle E,E^{\ast}_c\rangle$, where we say that a
subset $H$ of $E^{\ast}_c$ is ${\cal T}$-equicontinuous if $H$ is an equicontinuous
family of mappings from $(E,{\cal T})$ to
$(L^{0}({\cal F},K),|\cdot|_c)$.}

\th{Definition 6.2.3.} {\it Let $\langle E,F\rangle$ be a random
duality pair over $K$ with base $(\Omega,{\cal F},P)$ such that $F$
has the countable concatenation property. A family ${\cal B}$ of
$\sigma_c(F,E)$-bounded subsets of $F$ is called saturated if the
following are satisfied$:$

$(1)$ If $B\in{\cal B}$ and $A\subset B$, then $A\in{\cal B};$

$(2)$ If $A,B\in{\cal B}$, then $A\cup B\in{\cal B};$

$(3)$ If $B\in{\cal B}$, then $\sigma_c(F,E)-BC_4(B)\in{\cal B}$
$($see Definition $6.1.5$ for the operation $BC_4);$

$(4)$ If $B\in{\cal B}$ and $\lambda\in L^{0}({\cal F},K)$, then
$\lambda B\in {\cal B}$.}

\th{Proposition 6.2.4.} {\it Let $\langle E,F\rangle$ be a random
duality pair such that $F$ has the countable concatenation property.
If ${\cal B}$ is a saturated family of $\sigma_c(F,E)$-bounded
subsets of $F$, then ${\cal T}_{\cal B}$ is random admissible iff
$\cup{\cal B}=F$.}

\th{Definition 6.2.4 {\rm\cite{FKV}}.} {\it Let $(E,{\cal T})$ be a
locally $L^{0}$-convex module. A subset $G$ of $E$ is called an
$L^{0}$-barrel if it is $L^{0}$-convex, $L^{0}$-absorbent,
$L^{0}$-balanced and ${\cal T}$-closed. If every $L^{0}$-barrel is a
neighborhood of $0\in E$, then $(E,{\cal T})$ is called an
$L^{0}$-barreled module.}

\th{Definition 6.2.5.} {\it Let $(E,{\cal T})$ be a locally
$L^{0}$-convex module. If every $L^{0}$-barrel with the countable
concatenation property is a neighborhood of $0\in E$, then $(E,{\cal
T})$ is called an $L^{0}$-pre-barreled module.}\vspace{1mm}

Clearly, the two notions of an $L^{0}$-barreled module and an
$L^{0}$-pre-barreled module coincide for an ordinary locally convex
space, but the latter is weaker than the former in general. Up to
now, we have not yet known what kind of locally $L^{0}$-convex
module is $L^{0}$-barreled. Fortunately, we have the following:

\th{Proposition 6.2.5.}{\it Let $(E,{\cal T})$ be a locally
$L^{0}$-convex module such that $E$ has the countable concatenation
property. Then $E$ is an $L^{0}$-pre-barreled module iff ${\cal
T}=\beta(E,E^{\ast}_c)$, where $\beta(E,E^{\ast}_c)$ is the random
uniform convergence topology of $E$ over the family of all
$\sigma_c(E^{\ast},E)$-bounded subsets of $E^{\ast}_c$.}

Proposition 6.2.6 below is the most important result in this
section, since it is enough to meet the current needs of the theory
of conditional risk measures.

\th{Proposition 6.2.6.} {\it Every complete $RN$ module
$(E,\|\cdot\|)$ such that $E$ has the countable concatenation
property is an $L^{0}$-pre-barreled module when it is endowed with
the locally $L^{0}$-convex topology. Specially, $L^{p}_{\cal
F}({\cal E})$ is an $L^{0}$-pre-barreled module.}

\section{$L^{0}$-convex analysis and its applications to conditional risk measures}

The purpose of this section is to generalize the following three
basic theorems in classical convex analysis to random metric theory
and to apply the generalized basic theorems to the theory of
conditional risk measures.

To introduce the three basic theorems, let $E$ be a real Hausdorff
locally convex space and $E^{\prime}$ the classical conjugate space
of $E$. An extended real-valued convex function $f:E\rightarrow
[-\infty,+\infty]$ is called proper if $f(x)>-\infty$ for all $x\in
E$ and ${\rm dom}(f)=\{x\in E~|~f(x)<+\infty\}\neq\emptyset$.
Besides, ${\rm int}({\rm dom}(f))$ denotes the interior of $dom(f)$.

\th{Theorem A {\rm\cite{Ekeland-Temam}}.} {\it Every proper extended
real-valued lower semicontinuous convex function $f$ defined on a
barreled space $E$ is continuous on ${\rm int}({\rm dom}(f))$.}

\th{Theorem B {\rm\cite{Ekeland-Temam}}.} {\it Every proper extended
real-valued lower semicontinuous convex function $f$ defined on a
barreled space $E$ is subdifferentiable on ${\rm int}({\rm
dom}(f))$.}

\th{Theorem C {\rm \cite{Ekeland-Temam}}.} {\it For every proper
extended real-valued lower semicontinuous convex function $f$
defined on any locally convex space $E$, $f^{\ast\ast}=f$, where
$f^{\ast\ast}:E\rightarrow [-\infty,+\infty]$ is defined by
$f^{\ast\ast}(x)=\sup\{u(x)-f^{\ast}(u)~|~u\in E^{\prime}\},\forall
x\in E$ and $f^{\ast}:E^{\prime}\rightarrow  [-\infty,+\infty]$ is
defined by $f^{\ast}(u)=\sup\{u(x)-f(x)~|~x\in E\},\forall u\in
E^{\prime}$.}\vspace{1mm}

The so-called $L^{0}$-convex analysis is convex analysis of the
$L^{0}$-valued $L^{0}$-convex functions defined on random
locally convex modules.  $L^{0}$-convex analysis and its
applications to conditional risk measures were first studied by
Filipovi\'c et al. in \cite{FKV}, which is, without
doubt, an excellent contribution to both random metric theory and
the theory of conditional risk measures. On the other hand, there were
some negligences in their paper \cite{FKV}, their main results will
be improved based on Proposition 5.3 in the process of presenting
their results. Besides, to pave the way for applying random metric
theory to conditional risk measures, we further give the new
continuity and subdifferentiability theorems for $L^{0}$-convex
functions defined on $L^{0}$-pre-barreled modules because it is not
very convenient for the corresponding theorems given in \cite{FKV}
to be applied to conditional risk measures. In particular, we also
give a pleasant $(\varepsilon,\lambda)$-topological version of
Fenchel-Moreau type dual representation theorem for $L^{0}$-convex
functions, which contains the corresponding locally $L^{0}$-convex
topological version of Fenchel-Moreau type dual representation
theorem established in \cite{FKV} as a special case.

By the way, in this section the results without mention of a
reference belong to the author. To simplify the notation, let
$(\Omega,{\cal E},P)$ be a fixed probability space, $\cal F$ a fixed
sub $\sigma$-algebra of
${\cal E}$, we make the following convention:
$L^{0}({\cal E}):=L^{0}({\cal E},R)$,
$\bar{L}^{0}({\cal E}):=\bar{L}^{0}({\cal E},R)$,
$L^{p}({\cal E}):=\{\xi\in L^{0}({\cal E},R)~|~\int_{\Omega}|\xi|^{p}dP<+\infty\}$ for $1\leq p<+\infty$ and
$L^{\infty}({\cal E}):=\{\xi\in L^{0}({\cal E},R)~|~\xi$ is
essentially bounded$\}$.

Similarly, one can easily understand the notions such as
$L^{0}({\cal F}),\bar{L}^{0}({\cal F})$ and $L^{p}(\cal F)$ $(1\leq
p\leq+\infty)$.

\subsection{$L^{0}$-convex functions}

Let $E$ be a left module over the algebra $L^{0}(\cal F)$. The
effective domain of a function $f: E \rightarrow \bar{L}^{0}({\cal
F})$ is denoted by dom$(f):=\{x\in E~|~ f(x)\in L^{0}(\cal F)\}$.
The epigraph of $f$ is denoted by epi $(f):=\{(x,y)\in E\times
L^{0}(\mathcal{F})~|~f(x)\leq y\}$. The function $f$ is called
proper if $f(x)>-\infty$ on $\Omega$ for every $x\in E$ and
dom$(f)\neq \emptyset$.

\th{Definition 7.1.1 {\rm\cite{FKV,FKV-Approaches}}.} {\it Let $E$
be a left module over the algebra $L^{0}(\cal F)$ and $f:E
\rightarrow \bar {L}^{0}(\cal F)$.

$(1)$ $f$ is $L^{0}({\cal F})$-convex if $f(\xi x+(1-\xi)y)\leq \xi
f(x)+(1-\xi)f(y)$ for all $x$ and $y$ in $E$ and $\xi \in
L_{+}^{0}(\cal F)$ such that $0\leq \xi \leq 1$ (Here we make the
convention that $0\cdot (\pm \infty)=0$ and $\infty -\infty
=\infty)$.

$(2)$ $f$ has the local property if
$\tilde{I}_Af(x)=\tilde{I}_Af(\tilde{I}_{A}x)$ for all $x\in E$ and
$A\in \cal F$.

$(3)$ $f$ is regular if $\tilde{I}_Af(x)=f(\tilde{I}_{A}x)$ for all
$x\in E$ and $A\in \cal F$.}

\th{Proposition 7.1.1 {\rm\cite{FKV,FKV-Approaches}}.} {\it $f:E
\rightarrow \bar{L}^{0}(\cal F)$ is $L^{0}({\cal F})$-convex iff $f$
has the local property and epi$(f)$ is $L^{0}({\cal F})$-convex.}

\th{Definition 7.1.2.} {\it Let $(E,\cal F)$ be a random locally
convex module over $K$ with base $(\Omega,{\cal F},P)$. A function
$f:E \rightarrow L^{0}(\cal F)$ is called ${\cal
T}_{\epsilon,\lambda}$-continuous if it is continuous from $(E,{\cal
T}_{\epsilon,\lambda})$ to $(L^{0}({\cal F},K),{\cal
T}_{\epsilon,\lambda})$. A function $f:E\rightarrow L^{0}({\cal
F},K)$ is called ${\cal T}_{c}$-continuous if it is continuous from
$(E,{\cal T}_{c})$ to $(L^{0}({\cal F}, K),{\cal
T}_{c})$.}\vspace{1mm}

When is $L^{0}({\cal F})$-convex a function $:E\rightarrow
L^{0}(\cal F)$ if it is a convex (in the usual sense) function
defined on an $L^{0}(\cal F)$-module? We have the following pleasant
results:

\th{Proposition 7.1.2.} {\it Let $(E,\cal P)$ be a random locally
convex module over $R$ with base $(\Omega,{\cal F},P)$. Then a $\cal
T_{\epsilon,\lambda}$-continuous function $f:E\rightarrow L^{0}(\cal
F)$ is $L^{0}({\cal F})$-convex iff $f$ is convex and has the local
property.}\vspace{1mm}

Similarly, we can also obtain the following:

\th{Proposition 7.1.3.} {\it Let $f$ be a continuous function from
$L^{p}(\cal E)$ to $L^{r}(\cal F)$ $(1\leq p,r\leq +\infty)$. Then
$f$ is $L^{0}({\cal F})$-convex $($namely $f(\xi x+(1-\xi)y)\leq \xi
f(x)+(1-\xi)f(y)$, for all $x,y\in L^{p}(\cal E)$ and $\xi \in
L_{+}^{0}(\cal F)$ such that $0\leq\xi\leq 1$$)$ iff $f$ is convex
and has the local property $($namely
$\tilde{I_{A}}f(x)=\tilde{I}_Af(\tilde{I}_Ax)$ for all $x\in
L^{p}({\cal E})$ and $A\in \cal F$$)$.}

\subsection{Lower semicontinuity}

\th{Definition 7.2.1.} {\it Let $(E,\cal P)$ be a random locally
convex module over $R$ with base $(\Omega,{\cal F},P)$. A function
$f:E\rightarrow \bar{L}^{0}(\cal F)$ is called ${\cal
T}_{\epsilon,\lambda}$-lower semicontinuous if epi$(f)$ is closed in
$(E,\mathcal{ T}_{\epsilon,\lambda})\times(L^{0}(\cal F),\mathcal{
T}_{\epsilon,\lambda})$. A function $f:E\rightarrow \bar{L}^{0}(\cal
F)$ is called ${\cal T}_{c}$-lower semicontinuous if epi$(f)$ is
closed in $(E,{\cal T}_{c})\times (L^{0}({\cal F}),{\cal T}_{c})$.}

\th{Proposition 7.2.1.} {\it Let $(E,\cal P)$ be a random locally
convex module over $R$ with base $(\Omega,{\cal F},P)$ such that
both $E$ and $\cal P$ have the countable concatenation property. If
$f:E\rightarrow \bar{L}^{0}(\cal F)$ is a function with the local
property. Then the following are equivalent to each other$:$

$(1)$ $f$ is ${\cal T}_{c}$-lower semicontinuous$;$

$(2)$ $\{x\in E~|~ f(x)\leq r\}$ is ${\cal T}_{c}$-closed for each
$r\in L^{0}(\cal F);$

$(3)$ $\varliminf\limits_{\alpha} f(x_{\alpha})\geq f(x_{0})$ for
each $x_{0}\in E$ and each net $\{x_{\alpha},x\in \Gamma\}$ in $E$
such that $\{x_\alpha,\alpha \in \Gamma\}$ is ${\cal
T}_{c}$-convergent to $x_{0}$. Where
$\varliminf_{\alpha}f(x_{\alpha})=\vee_{\alpha \in
\Gamma}(\wedge_{\beta \geq \alpha}f(x_{\beta}))$.}

\th{Remark 7.2.1.} Proposition 7.2.1 first occurred in
\cite[Proposition 3.4 and Lemma 3.10]{FKV} where the countable
concatenation property of $E$ was not assumed, but this condition
should be added since Lemma 2.28 of \cite{FKV} has been improved to
Proposition 5.3.\vspace{1mm}

For the $(\epsilon,\lambda)$-topology, we only have the following:

\th{Proposition 7.2.2.} {\it Let $(E,\cal P)$ be a random locally
convex module over $R$ with base $(\Omega,{\cal F},P)$ and
$f:E\rightarrow \bar{L}^{0}(\cal F)$ a function. Then we have the
following statements$:$

$(1)$ $f$ is $\cal T_{\epsilon,\lambda}$-lower semicontinuous if
$\varliminf_{\alpha}f(x_{\alpha})\geq f(x_{0})$ for each $x_{0}\in
E$ and each net $\{x_{\alpha},\alpha \in \Gamma\}$ in $E$ such that
$\{x_{\alpha},\alpha \in \Gamma\}$ is $\cal
T_{\epsilon,\lambda}$-convergent to $x_{0}$;

$(2)$ $\{x\in E~|~ f(x)\leq r\}$ is $\cal
T_{\epsilon,\lambda}$-closed for each $r\in L^{0}(\cal F)$ if $f$ is
$\cal T_{\epsilon,\lambda}$- lower semicontinuous.}\vspace{1mm}

In general, a ${\cal T}_{\epsilon,\lambda}$-lower semicontinuous
function must be ${\cal T}_{c}$-lower semicontinuous. On the other
hand, Proposition 2.2.5 leads to the following nice result:

\th{Proposition 7.2.3.} {\it Let $(E,\cal P)$ be a random locally
convex module over $R$ with base $(\Omega,{\cal F},P)$ such that $E$
has the countable concatenation property and $f:E\rightarrow
\bar{L}^{0}(\cal F)$ a function with the local property. Then $f$ is
${\cal T}_{\epsilon,\lambda}$-lower semiconinuous iff $f$ is ${\cal
T}_{c}$-lower semicontinuous, specially this is true for an
$L^{0}({\cal F})$-convex function $f$.}

\subsection{Continuity and subdifferentiability}

Let $(E,\cal T)$ be a locally $L^{0}$-convex module over $R$
with base $(\Omega,{\cal F},P)$ and $f:E\rightarrow \bar{L}^{0}(\cal
F)$ a proper ${\cal T}$-lower semicontinuous $L^{0}({\cal
F})$-convex function. In this section, int(${\rm dom}(f$)) denotes
the $\cal {T}$-interior of ${\rm dom}(f)$. Further, $u\in E^\ast_c$
is called a subgradient of $f$ at $x_0\in {\rm dom}(f)$ if
$u(x-x_0)\leq f(x)-f(x_0),\forall x\in E$, and $\partial f(x_0)$
denotes the set of subgradients of $f$ at $x_0$. If $\partial
f(x_0)\neq\emptyset$, then $f$ is called ${\cal
T}$-subdifferentiable.

\th{Proposition 7.3.1 {\rm\cite{FKV}}.} {\it Let $(E,{\cal T})$ be
an $L^0$-barreled module over $R$ with base $(\Omega,{\cal F},P)$
and $f: E\rightarrow \bar {L}^0(\cal F)$ a proper $\cal {T}$-lower
semicontinuous $L^0$-convex function. Then $f$ is ${\cal
T}$-continuous on {\rm int}(${\rm dom}(f$)).}

\th{Proposition 7.3.2 {\rm\cite{FKV}}.} {\it Let $(E,{\cal T})$ and
$f$ be the same as in Proposition $7.3.1$. Then $\partial f(x_0)\neq
\emptyset$ for all $x_0\in$ {\rm int}$({\rm dom}(f))$.}

\th{Remark 7.3.1.} Proposition 7.3.2 is just Theorem 3.7 of
\cite{FKV} where $(E,{\cal T})$ was assumed to have the countable
concatenation property in the sense of \cite{FKV} (namely, ${\cal
T}$ can be a family $\cal P$ of $L^0$-seminorms on $E$ such that
$\cal P$ has the countable concatenation property), but the
assumption was not really used in the proof.\vspace{1mm}

In \cite{FKV}, Filipovic et al. stated in \cite[p.\,4018]{FKV} that
an $RN$ module would be $L^0$-barreled when it is endowed with the
locally $L^0$-convex topology. But the claim of them is obviously
not true, even it is not the case for a classical normed space,
either. What is more serious is that up to the present time we have
not known if $L_{\cal F}^p(\cal E)$ is an $L^0$-barreled module,
which considerably reduces the availability of Propositions 7.3.1
and 7.3.2 for conditional risk measures. Fortunately, the two
results below resolve the above difficulties!

\th{Proposition 7.3.3.} {\it Let $(E,{\cal T})$ be an
$L^0$-pre-barreled module over $R$ with base $(\Omega,{\cal F},P)$
such that $E$ has countable concatenation property and $f:
E\rightarrow \bar {L}^0(\cal F)$ a proper ${\cal T}$-lower
semicontinuous $L^0$-convex function. Then $f$ is ${\cal
T}$-continuous on {\rm int}$({\rm dom}(f))$.}

\th{Proposition 7.3.4.} {\it Let $(E,\cal T)$ and $f$ be the same as
in Proposition $7.3.3$. Then $\partial f(x_0)\neq \emptyset$ for all
$x_0$ in {\rm int}$({\rm dom}(f))$.}

\subsection{Fenchel-Moreau type dual representation theorems
under the two kinds of topologies}

Let $(E,\cal P)$ be a random locally convex module over $R$ with
base $(\Omega,{\cal F},P)$ and $f: E\rightarrow \bar {L}^0(\cal F)$
a proper ${\cal T}_{\epsilon,\lambda}$-lower semicontinuous
$L^0$-convex function.

The ${\cal T}_{\epsilon,\lambda}$-conjugate function
$f^\ast_{\epsilon,\lambda}: E^\ast_{\epsilon,\lambda}\rightarrow
\bar {L}^0(\cal F)$ of $f$ is defined as
follows:\[f^\ast_{\epsilon,\lambda}(u)=\vee \{u(x)-f(x)~|~x\in
E\},\quad \forall\,u\in E^\ast_{\epsilon,\lambda}.\]

The ${\cal T}_{\epsilon,\lambda}$-biconjugate function $f^{\ast
\ast}_{\epsilon,\lambda}: E\rightarrow \bar {L}^0(\cal F)$ of $f$ is
defined as follows:\[f^{\ast\ast}_{\epsilon,\lambda}(x)=\vee
\{u(x)-f^\ast_{\epsilon,\lambda}(u)~|~u\in
E^\ast_{\epsilon,\lambda}\},\quad \forall\,x\in E.\]

Then we have the $(\epsilon,\lambda)$-topological version of
Fenchel-Moreau type dual representation theorem as follows:

\th{Proposition 7.4.1.} {\it Let $(E,\cal P)$, $f$ and $f^{\ast
\ast}_{\epsilon,\lambda}$ be the same as above. Then $f^{\ast
\ast}_{\epsilon,\lambda}=f$.}\vspace{1mm}

Let $(E,\cal P)$ be a random locally convex module over $R$ with
base $(\Omega,{\cal F},P)$ and $f: E\rightarrow \bar {L}^0(\cal F)$
a proper ${\cal T}_c$-lower semicontinuous $L^0$-convex function.
Then the ${\cal T}_c$-conjugate function $f_c^\ast:
E^\ast_c\rightarrow \bar {L}^0(\cal F)$ of $f$ is defined by
$f^\ast_c(u)=\vee \{u(x)-f(x)~|~x\in E\}, \forall u\in E^\ast_c$.
And the ${\cal T}_c$-biconjugate function $f^{\ast \ast}_c:
E\rightarrow \bar {L}^0(\cal F)$ of $f$ is defined by:
$f^{\ast\ast}_c(x)=\vee \{u(x)-f^\ast_c(u)~|~u\in E^\ast_c\},
\forall x\in E$.

Then we can now have the ${\cal T}_c$-topological version of
Fenchel-Moreau type dual representation theorem as follows.

\th{Corollary 7.4.1.} {\it Let $(E,\cal P)$ be a random locally
convex module over $R$ with base $(\Omega,{\cal F},P)$ such that
both $E$ and $\cal P$ have the countable concatenation property and
$f$ a proper ${\cal T}_c$-lower semicontinuous $L^0$-convex
function. Then $f^{\ast \ast}_c=f$.}

\th{Remark 7.4.1.}Corollary 7.4.1 was first studied in \cite{FKV}
where the countable concatenation
property of $E$ was not assumed, but the condition should be added
to ensure the feasibility of the proof of \cite[Theorem 3.8]{FKV} as
given in \cite{FKV}. Since $E$ has  the countable concatenation
property, $f$ is proper, ${\cal T}_c$-lower semicontinuous and
$L^{0}({\cal F})$-convex iff $f$ is proper, ${\cal
T}_{\epsilon,\lambda}$-lower semicontinuous and $L^{0}({\cal
F})$-convex by Proposition 7.2.3, and when ${\cal P}$ has the
countable concatenation property $E^{\ast}_c$ is just
$E^{\ast}_{\epsilon,\lambda}$. Thus Corollary 7.4.1 is a special
case of Proposition 7.4.1, and Proposition 7.4.1 seems more natural
than Corollary 7.4.1 since Proposition 7.4.1 has the same form as
the classical Fenchel-Moreau type dual representation theorem
(namely Theorem C).

\subsection{Some applications to conditional risk measures}

\th{Definition 7.5.1 {\rm\cite{FKV-Approaches}}.} {\it Let $1\leq
p\leq+\infty$. A function $f:L^{p}_{\cal F}({\cal E})\rightarrow
\bar{L}^{0}({\cal F})$ is called$:$

$(1)$ monotone if $f(x)\leq f(y)$ for all $x,y\in L^{p}_{\cal
F}({\cal E})$ such that $x\geq y;$

$(2)$ subcash invariant if $f(x+y)\geq f(x)-y$ for all $x\in
L^{p}_{\cal F}({\cal E})$ and $y\in L^{0}_+({\cal F});$

$(3)$ cash invariant if $f(x+y)=f(x)-y$ for all $x\in L^{p}_{\cal
F}({\cal E})$ and $y\in L^{0}({\cal F});$

\noindent Further, an $L^{0}({\cal F})$-convex, monotone and cash
invariant function from $L^{p}_{\cal F}({\cal E})$ to
$\bar{L}^{0}({\cal F})$ is called an $L^{0}({\cal F})$-convex
conditional risk measure.}

\th{Remark 7.5.1.} An $L^{0}({\cal F})$-convex conditional risk
measure in the sense of Definition 7.5.1 is exactly a conditional
convex risk measure of $L^{p}_{\cal F}({\cal E})$-type as mentioned
in Section 1.3.\vspace{1mm}

Propositions 7.3.3 and 7.3.4 justify Proposition 7.5.1 below, which
was given in \cite{FKV}.

\th{Proposition 7.5.1.} {\it A proper ${\cal T}_c$-lower
semicontinuous $L^{0}({\cal F})$-convex function, in particular, a
proper ${\cal T}_c$-lower semicontinuous $L^{0}({\cal F})$-convex
conditional risk measure on $L^{p}_{\cal F}({\cal E})$, is ${\cal
T}_c$-continuous and subdifferentiable on the interior of its
effective domain.}\vspace{1mm}

Both Proposition 7.4.1 and Corollary 7.4.1 can justify Proposition
7.5.2 below that was first given in \cite{FKV}, but Proposition
7.4.1 seems more convenient for Proposition 7.5.3 below.

\th{Proposition 7.5.2.} {\it Let $1\leq p<+\infty$. Every proper
${\cal T}_c$ $($equivalently, ${\cal T}_{\epsilon,\lambda})$-lower
semicontinuous $L^{0}({\cal F})$-convex function $f:L^{p}_{\cal
F}({\cal E})\rightarrow \bar{L}^{0}({\cal F})$ can be represented as
follows$:$
$$f(x)=\vee\{E(x\cdot y~|~{\cal F})-f^{\ast}(y)~|~y\in L^{q}_{\cal F}({\cal E})\},\quad\forall\,x\in L^{p}_{\cal F}({\cal E}),$$
where $\frac{1}{p}+\frac{1}{q}=1$ and $f^{\ast}(y)=\vee\{E(x\cdot
y~|~{\cal F})-f(x)~|~x\in L^{p}_{\cal F}({\cal E})\},\forall y\in
L^{q}_{\cal F}({\cal E}).$}

\th{Proposition 7.5.3.} {\it Every proper
$\sigma_{\epsilon,\lambda}(L^{\infty}_{\cal F}({\cal E}),L^{1}_{\cal
F}({\cal E}))$ $($equivalently, $\sigma_{c}(L^{\infty}_{\cal
F}({\cal E}),L^{1}_{\cal F}({\cal E}))$$)$-lower semicontinuous
$L^{0}({\cal F})$-convex function $f:L^{\infty}_{\cal F}({\cal
E})\rightarrow\bar{L}^{0}(\cal F)$ can be represented as follows$:$
$$f(x)=\vee\{E(x\cdot y~|~{\cal F})-f^{\ast}(y)~|~y\in L^{1}_{\cal F}({\cal E})\},\quad\forall\,x\in L^{\infty}_{\cal F}({\cal E}),$$
where $f^{\ast}(y)=\vee\{E(x\cdot y~|~{\cal F})-f(x)~|~x\in
L^{\infty}_{\cal F}({\cal E})\},\forall y\in L^{1}_{\cal F}({\cal
E}).$}

\th{Remark 7.5.2.}Proposition 7.5.3 uses the fact that
$(L^{\infty}_{\cal F}({\cal E}),\sigma(L^{\infty}_{\cal F}({\cal
E}),L^{1}_{\cal F}({\cal E})))^{\ast}_{\epsilon,\lambda}=L^{1}_{\cal
F}({\cal E})$, which was proved in \cite{Guo-Chen}.\vspace{1mm}

When $f$ in Proposition 7.5.2 is a conditional risk measure, the
following refined Proposition 7.5.4 can be obtained:

\th{Proposition 7.5.4 {\rm\cite{FKV}}.} {\it Let $1\leq p<+\infty$.
Every proper ${\cal T}_c$ $($equivalently, ${\cal
T}_{\epsilon,\lambda})$-lower semicontinuous $L^{0}({\cal
F})$-convex conditional risk measure $f:L^{p}_{\cal F}({\cal
E})\rightarrow\bar{L}^{0}({\cal F})$ can be represented as
follows$:$
$$f(x)=\vee\{E(x\cdot y~|~{\cal F})-f^{\ast}(y)~|~y\in L^{q}_{\cal F}({\cal E}),y\leq 0\textmd{ and }E(y~|~{\cal F})=-1\},\quad\forall\, x\in L^{p}_{\cal F}({\cal E}),
$$
where $\frac{1}{p}+\frac{1}{q}=1$, and $f^{\ast}(y)$ is understood
as in Proposition $7.5.2.$}

\th{Proposition 7.5.5.} {\it When $p=\infty$, Proposition $7.5.4$ is
also valid if $f$ is a proper
$\sigma_{\epsilon,\lambda}(L^{\infty}_{\cal F}({\cal E}),L^{1}_{\cal
F}({\cal E}))$ $($or $\sigma_{c}(L^{\infty}_{\cal F}({\cal
E}),L^{1}_{\cal F}({\cal E}))$$)$-lower semicontinuous $L^{0}({\cal
F})$-convex conditional risk measure on $L^{\infty}_{\cal F}({\cal
E})$.}

\th{Example 7.5.1.}Let $\gamma>0$ and $1\leq p<+\infty$. Then
$\rho_{\gamma}:L^{p}_{\cal F}({\cal E})\rightarrow\bar{L}^{0}({\cal
F})$ defined by $\rho_{\gamma}(x)=\frac{1}{\gamma}\log E(e^{-\gamma
x}~|~{\cal F}),\forall x\in L^{p}_{\cal F}({\cal E})$, is a proper
${\cal T}_c$ (also ${\cal T}_{\epsilon,\lambda}$)$-$lower
semicontinuous $L^{0}({\cal F})$-convex conditional risk measure on
$L^{p}_{\cal F}({\cal E})$.\vspace{1mm}

Just as we have pointed out in Section 1.3, only the generalized Fenchel
Moreau type dual representation theorems\,---\,Proposition 7.4.1 and
Corollary 7.4.1, which are founded on the idea of random conjugate
spaces, can treat $\rho_{\gamma}$ thoroughly.

\section{Extensions of conditional risk measures}

In this section, we will prove that every conditional convex risk
measure of $L^{\infty}$-type which is representable as in Proposition
1.3.1 can be uniquely extended to a proper
$\sigma_{\epsilon,\lambda}(L^{\infty}_{\cal F}({\cal E}),
L^{1}_{\cal F}({\cal E}))$-lower semicontinuous conditional convex
risk measure of $L^{\infty}_{\cal F}({\cal E})$-type so that our
Proposition 7.5.5 implies Proposition 1.3.1. What is more important
is that we will also prove that every continuous convex conditional
risk measure of $L^{p}$-type can be uniquely extended to a ${\cal
T}_{\epsilon,\lambda}$-continuous conditional convex risk measure
from $L^{p}_{\cal F}({\cal E})$ to $L^{0}({\cal F})$ when $1\leq
p<+\infty$ so that Proposition 7.5.4 implies Proposition 1.3.3. Thus
the two representation Propositions 7.5.4 and 7.5.5 obtained along
the module approach unify all the previous representation
Propositions 1.3.3 and 1.3.1 obtained along the vector space
approach, respectively.

\th{Lemma 8.1.} {\it Let $f:L^{\infty}({\cal E})\rightarrow
L^{\infty}({\cal F})$ be a conditional convex risk measure of
$L^{\infty}$-type. Then there exists a unique ${\cal
T}_{\epsilon,\lambda}$-continuous conditional convex risk measure of
$L^{\infty}_{\cal F}({\cal E})$-type $\bar{f}:L^{\infty}_{\cal
F}({\cal E})\rightarrow L^{0}({\cal F})$ such that
$\bar{f}|_{L^{\infty}({\cal E})}=f$.}

\pf{Proof.}Let us first recall the definition of the $L^{0}$-norm
$|||\cdot|||_{\infty}:L^{\infty}_{\cal F}({\cal E})\rightarrow
L^{0}_+({\cal F})$ defined by $|||x|||_{\infty}=\wedge\{\xi\in
L^{0}_+({\cal F})~|~\xi\geq |x|\},\forall x\in L^{\infty}_{\cal
F}({\cal E})$. It is obvious that $|||x|||_{\infty}\in
L^{\infty}_+({\cal F})$ for any $x\in L^{\infty}({\cal E})$.

Since $x=y+x-y\leq y+|x-y|\leq y+|||x-y|||_{\infty}$ for any $x$ and
$y$ in $L^{\infty}({\cal E})$, we have that $f(x)\geq
f(y+|||x-y|||_{\infty})=f(y)-|||x-y|||_{\infty}$, namely
$f(x)-f(y)\geq -|||x-y|||_{\infty}$ and $f(y)-f(x)\leq
|||x-y|||_{\infty}$, so that $|f(x)-f(y)|\leq |||x-y|||_{\infty}$
for all $x$ and $y\in L^{\infty}({\cal E})$. Thus $f$ is uniformly ${\cal
T}_{\epsilon,\lambda}$-continuous from $(L^{\infty}({\cal
E}),|||\cdot|||_{\infty})$ to $(L^{\infty}({\cal F}),|\cdot|)$.
Further, since $L^{\infty}({\cal E})$ is ${\cal
T}_{\epsilon,\lambda}$-dense in $(L^{\infty}_{\cal F}({\cal
E}),|||\cdot|||_{\infty})$ by noticing $L^{\infty}({\cal
E})=L^{\infty}(L^{\infty}_{\cal F}({\cal E}))$ and making use of
Proposition 2.2.4, $f$ has a unique extension
$\bar{f}:L^{\infty}_{\cal F}({\cal E})\rightarrow L^{0}({\cal F})$
and it is easy to see that $\bar{f}$ is also a conditional convex
risk measure of $L^{\infty}_{\cal F}({\cal E})$-type.

Let ${\cal P}=\{Q~|~Q$ is a probability measure on $(\Omega,{\cal
E})$ such that $Q$ is absolutely continuous with respect to $P\}$
and ${\cal P}_{\cal F}=\{Q\in{\cal P}~|~Q=P$ on ${\cal F}\}$.
Further, we identify any element $Q$ in ${\cal P}$ with its
Radon-Nikod\'ym derivative $\frac{dQ}{dP}\in L^{1}({\cal E})$, then
${\cal P}_{\cal F}$ can be identified with the set $\{y\in
L^{1}_{+}({\cal E})~|~E(y~|~{\cal F})=1\}$, still denoted by ${\cal
P}_{\cal F}$, where $E(\cdot~|~{\cal F})$ denotes the conditional
expectation under the probability $P$. Then the random penalty
function $\alpha:{\cal P}_{\cal F}\rightarrow \bar{L}^{0}({\cal F})$
in Proposition 1.3.1 can be rewritten as
$\alpha(y)=\vee\{E(-xy~|~{\cal F})-f(x)~|~x\in L^{p}({\cal
E})\},\forall y\in {\cal P}_{\cal F}$; and if $f$ satisfies (1) of
Proposition 1.3.1 then $f(x)=\vee\{E(-xy~|~{\cal
F})-\alpha(y)~|~y\in {\cal P}_{\cal F}\}$. Define $f^{\ast}:\{y\in
L^{1}({\cal E})~|~y\leq 0, E(y~|~{\cal
F})=-1\}\rightarrow\bar{L}^{0}({\cal F})$ by
$f^{\ast}(y)=\alpha(-y)$, for any $y\in L^{1}({\cal E})$ such that
$y\leq 0$ and $E(y~|~{\cal F})=-1$, then $f(x)=\vee\{E(xy~|~{\cal
F})-f^{\ast}(y)~|~y\in L^{1}({\cal E}),y\leq 0$ and $E(y~|~{\cal
F})=-1\}$.

Finally, define $\bar{f}^{\ast}:L^{1}_{\cal F}({\cal
E})\rightarrow\bar{L}^{0}({\cal F})$ by
$\bar{f}^{\ast}(y)=\vee\{E(xy~|~{\cal F})-\bar{f}(x)~|~x\in
L^{\infty}_{\cal F}({\cal E})\},\forall y\in L^{1}_{\cal F}({\cal
E})$, where $\bar{f}$ denotes the unique extension of $f$ as
obtained in Lemma 8.1. Since $L^{\infty}({\cal E})$ is ${\cal
T}_{\epsilon,\lambda}$-dense in $L^{\infty}_{\cal F}({\cal E})$ and
$f$ is ${\cal T}_{\epsilon,\lambda}$-continuous, it is easy to see
that $\bar{f}^{\ast}(y)=\vee\{E(xy~|~{\cal F})-{f}(x)~|~x\in
L^{\infty}({\cal E})\},\forall y\in L^{1}_{\cal F}({\cal E})$, and
$\bar{f}^{\ast}(y)=f^{\ast}(y)$ for any $y\in L^{1}({\cal E})$ such
that $y\leq 0$ and $E(y~|~{\cal F})=-1$.

\th{Theorem 8.1.}\quad{\it Let $f:L^{\infty}({\cal E})\rightarrow
L^{\infty}({\cal F})$ be a conditional convex risk measure of
$L^{\infty}$-type. Then the following statements are equivalent to
each other$:$

$(1)$ $f(x)=\vee\{E(xy~|~{\cal F})-f^{\ast}(y)~|~y\in L^{1}({\cal
E}),y\leq 0$ and $E(y~|~{\cal F})=-1\},\forall x\in L^{\infty}({\cal
E});$

$(2)$ $f(x)=\vee\{E(xy~|~{\cal F})-\bar{f}^{\ast}(y)~|~y\in
L^{1}_{\cal F}({\cal E}),y\leq 0$ and $E(y~|~{\cal F})=-1\},\forall
x\in L^{\infty}({\cal E});$

$(3)$ $\bar{f}(x)=\vee\{E(xy~|~{\cal F})-\bar{f}^{\ast}(y)~|~y\in
L^{1}_{\cal F}({\cal E}),y\leq 0$ and $E(y~|~{\cal F})=-1\},\forall
x\in L^{\infty}_{\cal F}({\cal E});$

$(4)$ $\bar{f}$ is a $\sigma_{\epsilon,\lambda}(L^{\infty}_{\cal
F}({\cal E}), L^{1}_{\cal F}({\cal E}))$-lower semicontinuous
conditional convex risk measure of $L^{\infty}_{\cal F}({\cal
E})$-type from $L^{\infty}_{\cal F}({\cal E})$ to $L^{0}({\cal
F})$.}

\pf{Proof.}(1)$\Rightarrow$(2) is clear.

(2)$\Rightarrow$(1). We only need to prove that for each fixed $x$ in
$L^{\infty}({\cal E})$ and each fixed $y\in L^{1}_{\cal F}({\cal
E})$ such that $y\leq 0$ and $E(y~|~{\cal F})=-1, E(xy~|~{\cal
F})-\bar{f}^{\ast}(y)\leq\vee\{E(x\tilde{y}~|~{\cal
F})-f^{\ast}(\tilde{y})~|~\tilde{y}\in L^{1}({\cal E}),\tilde{y}\leq
0$ and $E(\tilde{y}~|~{\cal F})=-1\}$.

In fact, let $A_n=[E(|y|~|~{\cal F})\leq n]$ and $y_n=I_{A_n}\cdot
y+(1-I_{A_n})\cdot(-1)$ for each positive integer $n$, then $y_n\in
L^{1}({\cal E}), y_n\leq 0$ and $E(y_n~|~{\cal F})=-1$. Further,
since both $E(xz~|~{\cal F})$ and $\bar{f}^{\ast}(z)$ have the local
property with respect to $z$ in $L^{1}_{\cal F}({\cal E})$, then
$I_{A_n}(E(xy~|~{\cal F})-\bar{f}^{\ast}(y))=I_{A_n}E(x\cdot
I_{A_n}y~|~{\cal
F})-I_{A_n}\cdot\bar{f}^{\ast}(I_{A_n}y)=I_{A_n}(E(xy_n~|~{\cal
F})-\bar{f}^{\ast}(y_n))\leq I_{A_n}(\vee\{E(x\tilde{y}~|~{\cal
F})-f^{\ast}(\tilde{y})~|~\tilde{y}\in L^{1}({\cal E}),\tilde{y}\leq
0$ and $E(\tilde{y}~|~{\cal F})=-1\})$. Letting
$n\rightarrow\infty$, one can have that $E(xy~|~{\cal
F})-\bar{f}^{\ast}(y)\leq\vee\{E(x\tilde{y}~|~{\cal
F})-f^{\ast}(\tilde{y})~|~\tilde{y}\in L^{1}({\cal E}),\tilde{y}\leq
0$ and $E(\tilde{y}~|~{\cal F})=-1\}$.

(2)$\Rightarrow$(3). Define $g:L^{\infty}_{\cal F}({\cal
E})\rightarrow\bar{L}^{0}({\cal F})$ by $g(x)=\vee\{E(xy~|~{\cal
F})-\bar{f}^{\ast}(y)~|~y\in L^{1}_{\cal F}({\cal E}),y\leq 0$ and
$E(y~|~{\cal F})=-1)\},\forall x\in L^{\infty}_{\cal F}({\cal E})$,
then $g$ is ${\cal T}_{\epsilon,\lambda}$-lower semicontinuous. From
(2) one has that $g(x)=\bar{f}(x),\forall x\in L^{\infty}({\cal
E})$, so that $g(x)\geq \bar{f}(x),\forall x\in L^{\infty}_{\cal
F}({\cal E})$ since $\bar{f}$ is ${\cal
T}_{\epsilon,\lambda}$-continuous and $L^{\infty}({\cal E})$ is
${\cal T}_{\epsilon,\lambda}$-dense. Eventually,
$g(x)=\bar{f}(x),\forall x\in L^{\infty}_{\cal F}({\cal E})$, since
it is obvious that $g(x)\leq \bar{f}(x),\forall x\in
L^{\infty}_{\cal F}({\cal E})$.

(3)$\Rightarrow$(2) is clear.

(3)$\Rightarrow$(4) is clear.

(4)$\Rightarrow$(3) is implied by Proposition 7.5.5.

\th{Lemma 8.2.} {\it Let $1\leq r\leq p<+\infty$ and $f:L^{p}({\cal
E})\rightarrow L^{r}({\cal F})$ be a continuous convex conditional
risk measure of $L^{p}$-type. Then $f$ can be uniquely extended to a
${\cal T}_{\epsilon,\lambda}$-continuous conditional convex risk
measure of $L^{p}_{\cal F}({\cal E})$-type $\bar{f}$ from
$L^{p}_{\cal F}({\cal E})$ to $L^{0}({\cal F})$.}

\pf{Proof.}
We first prove that $f$ is ${\cal T}_{\epsilon,\lambda}$-continuous
from $(L^{p}({\cal E}),|||\cdot|||_p)$ to $(L^{r}({\cal
F}),|\cdot|)$, see Section 2.3 for the $L^{0}$-norm $|||\cdot|||_p$.
To this, we only need to prove that, for each fixed $x_0\in
L^{p}({\cal E})$ and each sequence $\{x_n,n\in N\}$ in $L^{p}({\cal
E})$ such that $\{E(|x_n-x_0|^{p}~|~{\cal F}),n\in N\}$ converges in
probability $P$ to 0, there exists a subsequence $\{x_{n_k},k\in
N\}$ of $\{x_n,n\in N\}$ such that $\{f(x_{n_k}),k\in N\}$ converges
in probability $P$ to $f(x_0)$. Since $f$ is monotone and cash
invariant, $f$ must be local, so that we only need to prove that,
for any positive number $\delta$, there exists an ${\cal
F}$-measurable subset $H_{\delta}$ of $\Omega$ and a subsequence
$\{x_{n_k},k\in N\}$ of $\{x_n,n\in N\}$ such that
$P(\Omega\backslash H_{\delta})>1-\delta$ and $\{f(x_{n_k}),k\in
N\}$ converges in probability $P$ to $f(x_0)$ on $\Omega\backslash
H_{\delta}$. In fact, by the Egoroff theorem  there are such
$H_{\delta}$ and $\{x_{n_k},k\in N\}$ such that
$\{E(|x_{n_k}-x_0|^{p}~|~{\cal F}),k\in N\}$ converges uniformly to
0 on $\Omega\backslash H_{\delta}$, so that
$\{\tilde{I}_{\Omega\backslash H_{\delta}}x_{n_k},k\in N\}$
converges to $\tilde{I}_{\Omega\backslash H_{\delta}}x_0$ in the
usual $L^{p}$-norm $\|\cdot\|_p$ by the Lebesgue convergence
theorem, hence $\{\tilde{I}_{\Omega\backslash
H_{\delta}}f(x_{n_k}),k\in N\}$ converges in the $L^{r}$-norm to
$\tilde{I}_{\Omega\backslash H_{\delta}}f(x_0)$, which implies that
$\{f(x_{n_k}),k\in N\}$ converges in probability $P$ to $f(x_0)$ on
$\Omega\backslash H_{\delta}$.

Next, it is easy to observe that $L^{p}_{\cal F}({\cal E})=\{\sum_{n=1}^{\infty}\tilde{I}_{A_n}x_n~|~\{A_n,n\in N\}\textmd{ is a countable partition of }\Omega\textmd{ to }{\cal F}$ and $\{x_n,n\in N\}\textmd{ is a sequence in }L^{p}({\cal E})\}$. Though $f$ is not necessarily uniformly ${\cal T}_{\epsilon,\lambda}$-continuous, the local property of a conditional
convex function motivates us to define $\bar{f}:L^{p}_{\cal F}({\cal E})\rightarrow L^{0}({\cal F})$ by
$$\bar{f}(x)=\sum_{n=1}^{\infty}\tilde{I}_{A_n}f(x_n),\forall x=\sum_{n=1}^{\infty}\tilde{I}_{A_n}x_n,$$
since $f$ is convex, local and continuous, $f$ must be $L^{0}({\cal
F})$-convex by Proposition 7.1.3, from which one can see that the definition of $\bar{f}(x)$ is independent of the expression of $x$ and it is not difficult to verify that $\bar{f}$ is a ${\cal T}_{\epsilon,\lambda}$-continuous conditional
convex risk measure of $L^{p}_{\cal F}({\cal E})$-type. Finally, Since $L^{p}({\cal E})$ is ${\cal T}_{\epsilon,\lambda}$-dense in $L^{p}_{\cal F}({\cal E})$, an $L^{p}_{\cal F}({\cal E})$-type of ${\cal T}_{\epsilon,\lambda}$-continuous conditional convex risk measure as an extension of $f$ must be unique.

\th{Lemma 8.3.} {\it Let $1\leq r\leq p<+\infty$ and $u:L^{p}({\cal
E})\rightarrow L^{r}({\cal F})$ be a continuous linear function with
the local property. Then $u$ can be uniquely extended to a  ${\cal
T}_{\epsilon,\lambda}$-continuous $L^{0}({\cal F})$-linear function
$\bar{u}$ from $L^{p}_{\cal F}({\cal E})$ to $L^{0}({\cal F})$ such
that the $L^{0}$-norm $\|\bar{u}\|$ of $\bar{u}$ satisfies
$\|\bar{u}\|\in L^{\frac{pr}{p-r}}({\cal F})$, where
$\frac{pr}{p-r}=\infty$ when $p=r$.}

\pf{Proof.}In the proof of Lemma 8.2 only the continuity and the
local property of $f$ are used for the existence of a ${\cal
T}_{\epsilon,\lambda}$-continuous extension $\bar{f}$, thus the same
reasoning shows that such a unique ${\cal
T}_{\epsilon,\lambda}$-continuous linear extension $\bar{u}$ with
the local property exists. Since $\bar{u}$ is linear and local,
$\bar{u}$ must be regular, namely
$\bar{u}(\tilde{I}_Ax)=\tilde{I}_A\bar{u}(x),\forall x\in
L^{p}_{\cal F}({\cal E})$ and $A\in{\cal F}$, hence $\bar{u}$ is
also $L^{0}({\cal F})$-linear by the ${\cal
T}_{\epsilon,\lambda}$-continuity of $\bar{u}$.

By Corollary 3.4, there exists a unique $y\in L^{q}_{\cal F}({\cal
E})$ such that $\bar{u}(x)=E(xy~|~{\cal F}),\forall x\in L^{p}_{\cal
F}({\cal E})$, where $q$ is the H\"older conjugate number of $p$. In
particular, $u(x)=E(xy~|~{\cal F}),\forall x\in L^{p}({\cal E})$,
then Proposition 2.5 of \cite{FKV-Approaches} shows that
$E(|y|^q~|~{\cal F})\in L^{\frac{r(p-1)}{p-r}}({\cal F})$, namely
$\|\bar{u}\|=E(|y|^q~|~{\cal F})^{1/q}\in L^{\frac{pr}{p-r}}({\cal
F})$.

Let $f$ and $\bar{f}$ be the same as in Lemma 8.2.

Define $f^{\ast}:\{y\in L^{q}({\cal E})~|~y\leq 0,E(|y|^q~|~{\cal
F})\in L^{\frac{r(p-1)}{p-r}}({\cal F})$ and $E(y~|~{\cal
F})=-1\}\rightarrow \bar{L}^{0}({\cal F})$ by
$f^{\ast}(y)=\vee\{E(xy~|~{\cal F})-f(x)~|~x\in L^{p}({\cal E})\}$.

Define $\bar{f}^{\ast}:L^{q}_{\cal F}({\cal E})\rightarrow
\bar{L}^{0}({\cal F})$ by $\bar{f}^{\ast}(y)=\vee\{E(xy~|~{\cal
F})-\bar{f}(x)~|~x\in L^{p}_{\cal F}({\cal E})\}$.

It is easy to see that $\bar{f}^{\ast}(y)=f^{\ast}(y),\forall y\in
L^{q}({\cal E})$ such that $y\leq 0, E(|y|^{q}~|~{\cal F})\in
L^{\frac{r(p-1)}{p-r}}({\cal F})$ and $E(y~|~{\cal F})=-1$.

\th{Theorem 8.2.} {\it Let $f$ and $\bar{f}$ be the same as in Lemma
$8.2$. Then the following statements are true and equivalent to each
other$:$

$(1)$ $f(x)=\vee\{E(xy~|~{\cal F})-f^{\ast}(y)~|~y\in L^{q}({\cal
E}),y\leq 0, E(|y|^{q}~|~{\cal F})\in L^{\frac{r(p-1)}{p-r}}({\cal
F})$ and $E(y~|~{\cal F})=-1\},\forall x\in L^{p}({\cal E});$

$(2)$ $f(x)=\vee\{E(xy~|~{\cal F})-\bar{f}^{\ast}(y)~|~y\in
L^{q}_{\cal F}({\cal E}),y\leq 0$ and $E(y~|~{\cal F})=-1\},\forall
x\in L^{p}({\cal E});$

$(3)$ $\bar{f}(x)=\vee\{E(xy~|~{\cal F})-\bar{f}^{\ast}(y)~|~y\in
L^{q}_{\cal F}({\cal E}),y\leq 0$ and $E(y~|~{\cal F})=-1\},\forall
x\in L^{p}_{\cal F}({\cal E})$.}

\pf{Proof.}(1) is exactly Proposition 1.3.3 by identifying $y$ in
(1) with $E(\cdot y~|~{\cal F})$ in Proposition 1.3.3.

(1)$\Rightarrow$(2) is clear, so that (2) is true.

(2)$\Rightarrow$(3) is similar to the proof of (2)$\Rightarrow$(3)
in Theorem 8.1, so that (3) is true (in fact, (3) can be obtained
from Proposition 7.5.4).

(3)$\Rightarrow$(2) is clear.

(2)$\Rightarrow$(1). Let $y\in L^{q}_{\cal F}({\cal E})$ be such that
$y\leq 0$ and $E(y~|~{\cal F})=-1$. For each positive integer $n$,
let $A_n=[E(|y|^{q}~|~{\cal F})\leq n]$ and
$y_n=I_{A_n}y+(1-I_{A_n})(-1)$, then $y_n\in L^{q}({\cal E}),
E(|y_n|^{q}~|~{\cal F})\in L^{\frac{r(p-1)}{p-r}}({\cal F}),y_n\leq
0$ and $E(y_n~|~{\cal F})=-1$, then similar to the proof of
(2)$\Rightarrow$(1) of Theorem 8.1 one can complete the remaining
part of the proof of (2)$\Rightarrow$(1).

To draw a conclusion, Lemmas 8.1 and 8.2 show that Definition 7.5.1
unifies both Definitions 1.3.1 and 1.3.2. Further, Theorems 8.1 and
8.2 show that Propositions 7.5.5 and 7.5.4 include Propositions
1.3.1 and 1.3.3 as a special case, respectively. In particular, only
the module approach to conditional risk measures based on Definition
7.5.1 can treat thoroughly conditional entropic risk measure
$\rho_{\gamma}$ together with many other conditional risk measures
as exhibited in \cite{FKV-Approaches}. Thus the module approach has
striking advantages. It should be expected that the deep development
of dynamic risk measures will involve more of random metric theory.

{\noindent\bf Acknowledgements} The author would like to thank Professor Shijian
Yan for some invaluable suggestions. When the author finished this paper, he was informed
of the three closely related references
\cite{Song-Yan,Song-Yan-risk,Song-Yan-survey} in communication with
Professor Jiaan Yan, the author would like to thank Professor Jiaan
Yan for providing these references which should have been cited in
Section 1.2. This work was supported by National Natural Science
Foundation of China (Grant No. 10871016).

\end{document}